\documentclass[aps,prd,nofootinbib,preprint]{revtex4}
\pdfoutput=1
\DeclareFontFamily{U}{rsfs}{}         
\DeclareFontShape{U}{rsfs}{m}{n}{<5> rsfs5 <6><7> rsfs7          %
  <8><9><10><10.95><12><14.4><17.28><20.74><24.88> rsfs10}{}     %
\DeclareMathAlphabet{\mathfs}{U}{rsfs}{m}{n}                     %
\usepackage[latin1]{inputenc}
\usepackage{mathrsfs}
\usepackage{amsmath,bbm}
\usepackage{amssymb}

\def\be{\nopagebreak[3]\begin{equation}}
\def\ee{\end{equation}}
\def\ba{\nopagebreak[3]\begin{eqnarray}}
\def\ea{\end{eqnarray}}

\newcommand{\teta}{\rlap{\lower2ex\hbox{$\,\tilde{}$}}\eta{}}

\newcounter{mnotecount}[section]

\begin{document}

\author{Alberto Diez-Tejedor}  
\email{alberto.diez@fisica.ugto.mx}
\affiliation{Instituto de Ciencias Nucleares, Universidad Nacional Aut\'onoma de M\'exico, Circuito Exterior C.U., A.P. 70-543,
  M\'exico D.F. 04510, M\'exico}
\affiliation{Departamento de F\'isica, Divisi\'on de Ciencias e Ingenier\'ias, Campus Le\'on, Universidad de Guanajuato, Le\'on 37150, M\'exico}

\author{Daniel Sudarsky}
\email{sudarsky@nucleares.unam.mx}
\affiliation{Instituto de Ciencias Nucleares, Universidad Nacional Aut\'onoma de M\'exico, Circuito Exterior C.U., A.P. 70-543,  M\'exico D.F. 04510, M\'exico}
\affiliation{Instituto de Astronom\'ia y F\'isica del Espacio, Casilla de Correos 67, Sucursal 28, Buenos Aires 1428, Argentina}

\date{ V.1 \today}

\title{Towards  a formal  description  of  the collapse  approach to  the inflationary origin of  the seeds  of cosmic  structure}

\begin{abstract}

Inflation plays a central role in our current understanding of the universe.
According to the standard viewpoint, the homogeneous and isotropic mode  of the  inflaton field 
drove an early phase of nearly exponential expansion of the universe, while the  quantum fluctuations ({\it uncertainties}) of 
 the other modes gave rise to the  seeds  of cosmic  structure.
However, if we accept that the  accelerated expansion led the universe into an essentially homogeneous and  
isotropic space-time, with the state of all the matter fields in their vacuum 
(except for the zero mode of the inflaton field), 
we can not escape the conclusion that the state of  the universe as a whole would remain always homogeneous and  isotropic. 
It was recently proposed in  [A. Perez,  H. Sahlmann  and D. Sudarsky, 
``On the quantum origin of the seeds of cosmic  structure," Class. Quant. Grav. \textbf{23}, 2317-2354 (2006)] that a collapse 
(representing physics  beyond the  established  paradigm, and presumably  associated  with a  quantum-gravity effect {\it \`a la} Penrose) of  
the state function of the inflaton field  might be the missing element, and thus would be  responsible  for the  emergence  of the
primordial inhomogeneities.  Here  we  will discuss a formalism that  relies  strongly on quantum field theory on curved  space-times, 
 and within  which we  can  implement  a detailed   description of such a process. 
The picture that  emerges clarifies many aspects of the problem, and  is  conceptually quite  transparent. 
 Nonetheless,  we  will find   that  the results  lead  us   to argue that  the resulting picture is \textit{not}  
fully  compatible with a purely   geometric description of space-time.

\end{abstract}

\maketitle

\section{Introduction}

Inflation represents an appealing  framework to understand the ``initial conditions" of the universe \cite{Peacock1998, 
Liddle:2000cg, Dodelson2003, Padmanabhan1993, Mukhanov2005, Weinberg2008, Lyth:2009zz}. The  early phase
 of accelerated expansion seems to resolve some of the classical  ``naturalness problems" of  the big bang model, such
  as the flatness, the horizon, the exotic relics or the entropy ones 
\cite{Guth:1980zm, Linde:1981mu, Albrecht:1982wi, Starobinsky:1980te}. 
Moreover, it is generally accepted that inflation  also  accounts for  the small  anisotropies observed in the cosmic microwave 
background (CMB), necessary for the subsequent  emergence  of cosmic structure
\cite{Hawking:1982cz, Guth:1982ec, Starobinsky:1982ee, Bardeen:1983qw, Mukhanov:1990me}. The accelerated expansion makes it possible 
to ``push" the short wavelength, causally connected  modes of the inflaton field to the largest scales of the observable universe. 
If, in addition, the accelerated expansion was (nearly) de-Sitter, and the state of the quantum fields  was a featureless Bunch-Davies 
``vacuum" (except for the zero mode of the inflaton field),
a single physical scale would appear in the   primordial spectrum 
of the quantum fluctuations (uncertainties): the  Hubble parameter. 
Roughly speaking, this is what lies behind  the appearance of 
a scale-free (Harrison-Zel'dovich) primordial power
 spectrum of density perturbations in the standard model of inflation. 

However,  upon  a deeper examination, one finds that there is something 
rather strange  in that picture: if the observable
universe starts from the Bunch-Davies vacuum for the quantum  fields in a 
Robertson-Walker space-time (both of which are completely  homogeneous and isotropic), 
and considering their evolution according to the rules of standard physics, the state  of the quantum fields, and the universe 
as a whole, will remain both homogeneous and isotropic at all times. But, how can this be compatible with 
the highly complex structure of the universe we are inhabiting? In fact, how can such proposal be said to explain the emergence
 of the seeds of cosmic  structure,
which  early traces we  study  in the CMB? (See the references \cite{Perez:2005gh, Sudarsky:2009za}  for a full discussion 
of the problem  and the various attempts to deal  with these issues  within the standard physical paradigms.)
The standard   and  phenomenologically  successful  accounts on this  matter rely on the identification of the \textit{quantum} fluctuations  of  certain 
 observables associated to a homogeneous and isotropic universe, with the averages over an ensemble of inhomogeneous 
 universes of their analogue \textit{classical} quantities: more specifically  (or more crudely), on the identification
  of \textit{quantum uncertainties} and \textit{classical perturbations}. 
 As discussed  in  \cite{Sudarsky:2009za},  one  can not  avoid  concluding that
using standard physics there is no justification for such kind of identification. 
In fact, these shortcomings  have  started  to be recognized   by others in the literature: see for instance Section 10.4 in \cite{Padmanabhan1993}, the end of Section 8.3.3 in \cite{Mukhanov2005}, Section 10.1 in \cite{Weinberg2008}, Section 24.2 in \cite{Lyth:2009zz}, and Section 30.14 in \cite{Penrose2005}.

On the other hand, it is undeniable that the standard approach has a great phenomenological success: its  ``predictions" 
 match  exquisitely well  the most accurate observations  to date.
Nonetheless, as it has been argued in \cite{Perez:2005gh, Sudarsky:2009za}, 
 some new element must be  added  to the standard picture in order to have a physical  
 description of the transition from the initial symmetric stage 
 of the universe to the late  non-symmetric one, preserving at  the same time the phenomenological  success of the usual  approach.

In this  context,  it was recently proposed 
that a  new  kind  of  effect, which  at the phenomenological level  would be analogous to a ``self-induced" collapse 
in the state function of the inflaton field, could play  the required  role in the early universe \cite{Perez:2005gh} 
(see also \cite{Sudarsky:2007tp, Sudarsky:2006zx, DeUnanue:2008fw, Leon:2010fi, Leon:2010wv, DiezTejedor:2011jq}).
The point, of course, is that such kind of  collapses are a  serious departure form the unitary evolution of standard physics. 
It is  worth mentioning that, in fact, conceptually  similar   ideas have been considered  by
 many physicists concerned  with the measurement  problem in quantum mechanics 
(see for instance \cite{Bassi:2003gd} and references therein), and  the proposal that some quantum
  aspects associated to the gravitational interaction might  lie  at the root  of  this effect only  serves to 
make  the  idea  even more attractive \cite{Penrose:1996cv} (see also Chapter 30 in reference \cite{Penrose2005}). 
The  suggestion  that such a  mechanism could play a determinant role 
in the breakdown of the initial symmetry of the universe 
 seems  doubly appealing to us: On the one hand,  as we  will see, it  turns out to  have  the features leading to the kind of  effect    
  required for the emergence of the seeds of cosmic structure. On the other,  when we take such  collapse  as a  fundamental ingredient  
 for the  generation of primordial perturbations, these  become an   
 observationally accessible stage where  such a novel aspect of physics,  
 presumably containing some hints about the  nature of gravity at the quantum level,  
  could be  investigated.  
 Interestingly enough, some of the resulting   predictions  seem to differ in
  various aspects from the standard  accounts \cite{Perez:2005gh, DiezTejedor:2011jq}.
  The  main objective of this  work  is to  establish a formalism  that  can   serve as a basis  to explore these  ideas  in a well
  defined and  rigorous  setting,   which  would,  in  the future,  allow  us to  address  questions  such as  the applicability of  
collapse theories \cite{Bassi:2003gd, Karolyhazy1966, Diosi1984, Diosi1987, Diosi1989,GRW1,Pearle1989} to the inflationary  
universe,   and  to unearth  the  problems that  might arise in each  particular proposal. Our program is  based  on  the  joint use  
of quantum  field theory in curved  space-time and semiclassical gravity
  in  self-consistent settings.  However, as we  will see,  we  will  need to introduce a  modification/addition to that  general  setting, in connection with the  hypothetical collapse  of  the wave function.
  We  will describe all  that in detail in  the main parts of the  manuscript.

At this point we  should note that inflation is expected to be  associated with  very high energy scales,  well beyond the regime
 of the physical processes we have   otherwise explored. Nonetheless, it is  generally  expected that  such  a regime is not  
 really close to those where a quantum description of the gravitational interaction  would be required (see for instance Chapter 14 in reference \cite{Wald}). However, the fact that 
 we are interested in a situation  where the matter fields affecting the  space-time  geometry require a quantum description
  suggests that the appearance of   some effects,   ultimately traceable to a theory of quantum gravity,  should not be  too surprising. 
  In fact,   we should  keep in mind that inflation, to the extent that
  it is observationally accessible  through  its imprints in the  seeds of cosmic structure, represents 
  the  \textit{unique}  situation available to  us which requires   a treatment  which  simultaneously relies on
   both quantum theory   and Einstein's  gravity. 
  (There are several experiments which people often claim are  
   related to the quantum/gravity interface,  such as  the  famous  neutron  interference  experiment \cite{Colella1975} suitable 
   for detecting the effects of the Earth's  Newtonian  potential. However,  a detailed analysis \cite{Chryssomalakos:2002bm} 
   indicates that  these  experiments  are not sensitive to curvature at all, and it is  curvature where  gravitation truly  
   lies,  according to our post-Einsteinian views.)

 The paper is organized  as follows. In Section \ref{Effective} we review the ideas 
 regarding the collapse
  of the state function, attempting to frame the proposal, even if only 
  schematically, within our current understanding of physical theory. 
  We also introduce  
  some useful concepts to deal with the physics of the collapse,
  describing the nature of the formalism to be employed and explaining 
   the extent  to which it is taken to  describe the   relevant phenomena.
In Section \ref{formalizing} we
 study a  simplified  version of a wave function  collapse in the 
early universe,  describing 
the extent to  which the  transition from the symmetric to
 the inhomogeneous phase can be treated  within the present formalism. 
Finally, we discuss  our results in Section \ref{conclusions}.
We have placed in  the appendices some details of the presentation which are not essential
to follow the main text.

The  conventions  we will be using include:  $(-,+,+,+)$
 signature for the space-time metric,  Wald's convention for the Riemann tensor, and natural units with $c=1$. 
The Newton's gravitational constant $G$ and the reduced Planck's constant $\hbar$ are used to define 
the Planck mass $M_p^2\equiv\hbar^2/8\pi G$ and the Planck time $t_p^2 = 8\pi G\hbar$.
Space-time indexes are denoted by Greek letters ($\mu=0,1,2,3$). We use Latin letters for spatial indexes ($i=1,2,3$) associated  with  specific  hypersurfaces.

\section{The nature of our effective description}\label{Effective}
  
Before we  embark on the description  and the  general  setting for our  ideas, we  would like to remind  the reader that, despite the claims to the contrary, the measurement problem
 is still an open question in quantum mechanics \cite{Bassi:2003gd}. 
In fact, the problem  becomes  even   more   serious  
in  the context of   quantum field theory \cite{Sorkin:1993gg}. 
Even worse, the interpretational framework of the quantum theory becomes
 intolerable when we  want to apply it to the universe as a whole, 
 because, in that case,  we  can not even rely on the practical usage of  identifying an observer  and/or  a measuring  device. 
 Those  problems  have led  researchers,  such as  J. Hartle,  to argue  for  a  modified or extended   interpretational   framework  
  making the  theory  suitable  for the cosmological applications \cite{Hartle:1997hw, Hartle:2005ie}. However, it seems that  such  scheme
    does not  go far enough, lacking  the kind  of feature  we  need to address the  problem that concern us here \cite{Sudarsky:2009za}. 
    It is for that reason that we  shall  reconsider in some  detail the  schematic view
      proposed in \cite{Perez:2005gh} for dealing  with   this  conundrum.
 As we have just commented in the Introduction,  the proposal involves  incorporating 
 a ``self induced collapse of the wave function of the matter fields", 
and  considering the resulting inhomogeneities  
and anisotropies in the  energy momentum tensor generated during the collapses as the  origin for the   inhomogeneities  
and anisotropies  of  the space-time metric of our universe.


Now let us characterize the general nature of our approach. 
We should start by offering a discussion of a plausible manner whereby  
the collapse hypothesis might be incorporated within the 
general current understanding of a physical theory. However, for conciseness 
of the presentation this is done in the Appendix \ref{A}. 
Following the  general arguments given there, 
we will take  the  view  that the fundamental  degrees of freedom for the gravitational interaction 
 are not  the ones characterizing the standard  metric variables,  
but  some other variables that  are indirectly connected to those 
(consider,  for  example, 
the  fluxes and holonomies in loop quantum gravity \cite{Rovelli}).
 That is, we  will consider that  geometry is an emergent phenomena, and that general relativity should  be 
  regarded as some  sort of  \textit{hydrodynamical} limit of a more fundamental theory.\footnote{Hydrodynamics is
   the effective theory describing the long wavelength, large frequency, low energy modes of highly 
   coupled systems with a large number of ``particles". At those scales, only a small number of degrees
    of freedom are necessary to characterize the state of the system: {\it the hydrodynamic modes}. 
    In the case of a perfect fluid, it can be described by the energy density, the entropy density and the
     field of velocities; all related by the Euler, the continuity and the thermodynamic equations. (Notice that it has not been necessary to
      introduce the concept of a fundamental constituent $-$the particles$-$ for the fluid,  of which, the $18^{th}$ century  
 physicist who developed the theory was naturally  unaware.)
This description will be appropriate to describe the fluid at length scales larger than the mean free path, and for time
scales larger than the mean free time.
The view  we  will take here is that  something similar occurs  with   the gravitational interaction, where 
the characteristic physical scale should be probably determined now by the Planck mass. That  is, in fact, the situation in  
various  approaches to quantum gravity, such  as  the loop quantum gravity  program  or the  poset proposal. } 
This point of view seems to be  further supported by the analogy 
between the behaviour of black holes and the laws of thermodynamics \cite{Bekenstein:1973ur},
and by the ideas developed in \cite{Jacobson:1995ab} (see also \cite{Volovik2003, Padmanabhan:2003gd, Hu:2005ub}).

The  regimes  where the metric  description  
for the   gravitational degrees of freedom  
holds would  correspond to the situations where the whole universe might be described by states 
 $\vert\xi\rangle=\vert\xi\rangle_{g}\otimes\vert\xi\rangle_{m}$ (where the first  factor  corresponds to the  gravitational degrees of freedom, 
 and  the second one to the matter fields) for which the  gravitational sector   $\vert\xi\rangle_{g}$ is such that operators  corresponding to the metric  
  variables  have   sharply peaked values
 (this  would not be  a strict eigenstate of the metric, where   
the wave function would be a delta function, but only to states where the  wave function uncertainties in the geometrical  
quantities are rather small at the level of precision one is working).\footnote{Continuing with our hydrodynamical analogy, we  might  consider the very complicated
   wave-function characterizing the state of all the particles in a perfect fluid, but  which for practical purposes can be effectively
    described  by  some  classical  quantities  such as  the energy density, the entropy density, and the field of velocities. 
    The components of the space-time metric play now the role of the hydrodynamic modes.}
Needless is to say  that all our existing experience with the gravitational interaction would be, according to this view, well described by such hydrodynamical limit, 
and its non-geometrical behaviour would show up,  generically,  just in the extreme  domains  such as those  normally  associated  with  quantum  gravity.
Well outside those elusive  domains,
the  fundamental operators for  the  gravitational  degrees of freedom  would be,  for the most part,  well  described in terms
 of the metric  tensor  $ g_{\mu\nu}$, and its dynamics can be taken as controlled  by the corresponding Einstein  tensor
$ {}_{g}\langle\xi \vert \hat G \vert\xi\rangle_{g}=   G_{\mu\nu} [ g] $. At the same time, we will consider that the matter fields should be 
described in terms of a quantum field theory in a curved space-time. 
 Under these conditions 
 one can  expect to recover  the  semiclassical description of gravitation in interaction 
 with quantum  fields as reflected  in the semiclassical Einstein field
equations
\begin{equation}
\label{EE}
 G_{\mu\nu} [ g] =8\pi G\, {}_{m}\langle\xi\vert \hat{T}_{\mu\nu}[g]\vert\xi  \rangle_{m} .
 \end{equation}
Here the right hand  side  stands for the  expectation value of the energy-momentum tensor operator in the corresponding
  state of the quantum fields,  constructed from the  field operators  and the
space-time  metric  $ g_{\mu\nu}$. The matter fields are 
described in terms of operators acting 
on an appropriate Hilbert space $\mathscr{H}_m$,  according to the
standard construction of a quantum field theory  
on a background space-time,
with  the latter satisfying the semiclassical equations given in (\ref{EE}).

Given the  fact  that we do not have yet a  fully  workable quantum theory for the gravitational interaction, it is very difficult to analyse explicitly  the 
regime of validity of this semiclassical description.  However, it seems reasonable to assume that it would be appropriate for the situations in which  
the matter fields $\vert\xi\rangle_m$ are sharply peeked around a classical field configuration $(\varphi_{i,\xi}(x),\pi_{i,\xi}(x))$,
 with $\varphi_{i,\xi}(x)\equiv\, {}_{m} \langle\xi\vert\hat{\varphi}_i(x)\vert\xi\rangle_{m} $ 
 and $\pi_{i,\xi }(x)\equiv\,  {}_{m}\langle\xi\vert\hat{\pi}_i(x)\vert\xi\rangle_{m} $,  and  where there are no relevant  Planck  scale phenomena, 
i.e.  all curvature scalars  are  well  below  the Planck  scale.
%
%
Furthermore, we will be considering that the regime
of validity of semiclassical gravity, with  a relatively  simple  modification,
  can  be  extended to include the  self-induced  collapse  (or dynamical reduction) of the wave function of matter fields. Our goal is to  consider a   precise formalism  
  able to incorporate  these ideas, that will allow  us to  explore them in a  detailed manner.
It is worthwhile noting here  that  the  general  formalism to be  presented in the  first part of the  present work
 should be useful for the detailed 
 studies  of the dynamical wave function collapse
 theories (such as \cite{Bassi:2003gd, Diosi1984, Diosi1987, Diosi1989, GRW1, Pearle1989}) 
in the situations where the gravitational back-reaction
becomes important, as well as for certain aspects encountered  in related  approaches. 
   Each one of the discontinuous modifications  of  the evolution  equations, which are  controlled  by
  stochastic  functions  introduced  in   those  schemes,  would  correspond  to  
one single collapse as the one we will describe in this paper. 
  The  details of such  correspondence  would  clearly  be  different  for  each   
specific case. 
As an example, we consider briefly the so called stochastic gravity 
proposal \cite{Stoch} in Appendix \ref{B}.

Next we turn to the development of the precise  formalism  we  seek.
The description of the system in the periods between collapses is considered 
in Section \ref{SSCs}. In Section \ref{beyond the SSC} we describe  the proposal 
for incorporating the collapses  within the  general formalism.

 \subsection{The Semiclassical Self-consistent Configurations (SSCs)}\label{SSCs}
 
It  is clear that the semiclassical regime that we   have described  so  far  can not be  but an effective  description with a limited  range of  applicability.  
However,  we  will assume that  such regime includes the cosmological  setting at hand.
Next, we seek  to  specify  in some detail  the nature of the  formal description we will be considering,  even
  though we do not take it to  be,  in any sense,  fundamental.  
Among the advantages of having such a precise  structure,
 is that it allows  us to uncover and,  in principle,  to start  investigating, the places  where a departure from such  scheme  is in fact 
  required. As we  will see, this approach  will allow us to discuss 
 with precision some delicate questions, and to focus sharply  on some of  the 
less understood aspects of the  collapse ideas.
  (Consider for comparison  the formal developments in classical general relativity, which include the famous singularity theorems, 
   which as we know  indicate a limitation of the validity of that very same theory.)  

 Under these assumptions (and  at the above  specified level) we will consider that the universe 
can be  described  by what we call a  \textit{Semiclassical Self-consistent Configuration} (SSC). 
That is,  a space-time  geometry characterized by a  classical  space-time  metric  and  a standard quantum
 field theory  constructed  on that  fixed space-time background,
together with a particular  state in that construction
such that the  semiclassical   Einstein  equations 
hold.  
In other words, we will say that 
the set $\left\lbrace g_{\mu\nu}(x),\hat{\varphi}(x),\hat{\pi}(x),\mathscr{H},\vert\xi\rangle\in\mathscr{H}\right\rbrace$
represents a 
SSC if and only if $\hat{\varphi}(x)$, $\hat{\pi}(x)$ and $\mathscr{H}$  correspond to  a quantum field theory  
constructed over a space-time  with  metric  $g_{\mu\nu}(x)$ (as  described in, say \cite{Wald1994}), and  the  state 
$\vert\xi\rangle$ in $\mathscr{H}$  is  such that
\begin{equation}\label{Mset-up}
G_{\mu\nu}[g(x)]=8\pi G\langle\xi\vert \hat{T}_{\mu\nu}[g(x),\hat{\varphi}(x),\hat{\pi}(x)]\vert\xi\rangle 
\end{equation}
for all the points $x
$ in the space-time manifold. 
That  is,  we  are basically relying  on a strict interpretation of  semiclassical  gravity,
 considered not as  a fundamental theory,  but as  an effective  description
(Note  here  that,  as discussed in reference \cite{Carlip2008}, it  might even  
be  possible to take the alternative  
viewpoint.) It is  worth noting that  an  analogous  approach is  sometimes used 
in the field  of quantum optics  when 
 one  is not interested in some intrinsically quantum aspects of the  electromagnetic field 
(see for instance the ``self-consistent equations" in Section 1.5 of reference 
 \cite{Scully}).
We should  keep in mind   the self-referential feature of this approach,  and  also  the fact that  it is  very  close, in  spirit, to the   Schr\"odinger-Newton  
description \cite{Collapse-N-Sch1, Collapse-N-Sch2, Collapse-N-Sch3, Collapse-N-Sch4, Collapse-N-Sch5, Collapse-N-Sch6}, 
which  seems to have  a relationship  with some  kind of  collapse-like behaviour  \cite{Diosi1984, Sch-N}.
From now on we will ignore the subindexes ``$m$" and ``$g$", as it would be  
understood that the quantum description refers only to  the matter fields.
For simplicity, a single scalar field $\varphi$ has been assumed, where 
the evident generalization is understood.

The  actual   construction of  even one  of such  SSC's   is not  trivial. One must  somehow  ``guess"  the appropriate space-time metric, 
 construct  the quantum theory for the matter fields ``living" on that given background, and then, find  an appropriate 
state $\vert\xi\rangle$ in $\mathscr{H}$ (if any) compatible with the selected space-time configuration. 
We should emphasize that, in general, for a given SSC, most of  the states in $\mathscr{H}$ together with  the original 
$g(x)$, $\hat{\varphi}(x)$, $\hat{\pi}(x)$ and $\mathscr{H}$  do {\it not}  represent a valid  SSC. 
Only for a few states, if any (besides the  original one), will the  equation (\ref{Mset-up}) hold.
Explicit constructions of two SSCs will be  given in Sections \ref{before the collapse} and \ref{After the collapse}. The first one  
corresponds to a perfectly homogeneous and isotropic 
universe, and the second to a slightly inhomogeneous space-time, where,  for simplicity, just one Fourier component is taken to  be ``excited".

In order to avoid possible future misunderstandings, we should note at this point  various important differences between this 
formalism and the one usually employed  within the context of the inflationary universe.  The standard  approach  is based on
 the consideration of a  classical   description for both,  the space-time metric (taken to
  admit a flat Robertson-Walker  description) and inflaton 
 field (taken to be in a slow-rolling homogeneous and isotropic  configuration) ``backgrounds'', 
 and the  perturbations of both  metric and  inflaton field,  which 
 are treated at the quantum mechanical level.
In contrast, in the present formalism,   the split  between the  quantum and  classical descriptions  is  not tied to  a particular 
perturbative  approach, but to the gravity-matter distinction.
This seems to be  justified, as we have  argued  above, not only by the lack of a workable  theory of  quantum gravity, 
 but also by the conceptual problems that  seem  intrinsic  to that program, and, in  particular, to the so called  ``problem of time in quantum gravity''.  
One might be concerned with the fact that it seems always possible to move part 
of the degrees of freedom from the metric to the matter fields (and back) 
through a conformal transformation, and with the idea that changes of coordinates
mix the gravity and matter field perturbations.
These are common misunderstandings, and to their clarification we have devoted the Appendix \ref{app}.

 \subsection{Beyond a single SSC: the collapse}\label{beyond the SSC}
 
It should be  clear that  the Semiclassical Self-consistent Configurations introduced in the previous subsection are not enough in order to deal with the problem at hand, i.e. the transmutation 
of a symmetric universe into the actual inhomogeneous and anisotropic one. The extra element we must  consider in order to represent
our ideas is  the quantum  collapse of the   wave function of the 
matter fields. 
That is, the proposal that  the 
normal unitary evolution characterizing the standard field theory (and then by definition the SSCs) should be 
supplemented  by  instances of  quantum collapse,
 thought to   be triggered,  somehow,  by the  effects  of the gravitational degrees  of freedom which are not fully represented
  in the metric description (as it was indicated, the metric tensor should be   regarded here as  a mere 
  effective description $-$at the hydrodynamical level$-$  of the average 
  aggregate behaviour of the true  gravitational degrees of freedom).  
One of the main purposes of this paper is to propose a formalism that would  allow  us 
 to represent and  explore such idea, and to uncover its limitations.\footnote{Note 
that, despite the previously mentioned indications of collapse-like behaviour  
in the related  Schr\"odinger-Newton system, we are postulating the collapse  as   an additional feature, because  
it does not  seem that  those  previously  observed features  are  able to  
account  for the breaking of the initial translational and  rotational 
symmetries of the configuration.  That is,  the  Shr\"odinger-Newton system,  
once  provided  with an initial data  possessing one  such  symmetry,  would result  in  an equally  symmetric solution,  simply due to  the  deterministic  nature of the  problem and the fact that the dynamics does not break  those symmetries.}
 
We will separate the evolution of the system into the standard part and the collapse,  
considering the first one as described in terms of the Heisenberg picture, while the collapse will be treated in 
terms of the Schr\"odinger one. We  can look at that distinction as an ``interaction picture description", 
where  the role of the interaction is played by whatever physics lies behind the collapse process, and the rest is absorbed 
into the evolution of the operators, as it is done in the Heisenberg picture. 
Then, the space-time dependence encoded  in  the field  operators  reflects the standard  unitary evolution, where  
the  states will remain constant  except  when  a collapse occurs,  which will be characterized by  a  random jump  
of the state  $ |\xi \rangle $  to  one  among a  set of suitable   related  states $\lbrace |\zeta_1 \rangle ,\ldots,|\zeta_n \rangle\rbrace$,
in  what  we  will call   a ``self-induced"  collapse of the  wave  function, 
\begin{equation}
\label{Col}
|\xi \rangle   \to |\zeta \rangle.
\end{equation}
However, recall that, as we argued in Section \ref{SSCs}, any state  involved in 
the  characterization of the  conditions in the universe  should be 
always understood within a particular SSC. It is for that reason that it  is more appropriate to re-express the notion depicted in  
(\ref{Col}) in the more precise form
\begin{equation}
\label{}
\textrm{SSC}\textrm{-I}   \to \textrm{SSC}\textrm{-II}.
\end{equation}

In order to proceed,  we need  a  rather precise prescription for the  description of ``the collapse".
Consider first, within the Hilbert space associated to the  given SSC-I, that a transition $\vert\xi^{\textrm{(I)}}\rangle\to\vert\zeta^{\textrm{(I)}}\rangle_{\textrm{target}}$ 
``is about to happen", with both $\vert\xi^{\textrm{(I)}}\rangle$ and $\vert\zeta^{\textrm{(I)}}\rangle_{\textrm{target}}$ in $\mathscr{H}^{\textrm{(I)}}$.
Generically, the set $\{g^{\textrm{(I)}},\hat{\varphi}^{\textrm{(I)}},\hat{\pi}^{\textrm{(I)}}, \mathscr{H}^{\textrm{(I)}},\vert\zeta^{\textrm{(I)}}\rangle_{\textrm{target}}\}$ will 
not represent a new SSC.  We  will thus  say that the state $ \vert\zeta^{\textrm{(I)}}\rangle_{\textrm{target}}$ is ``not physical".  It represents  a    characterization  
(of sorts) of the state into which the collapse  will  take our  matter fields, 
employing the mathematical  language of the $\mathscr{H}^{\textrm{(I)}}$, a  language that would  be  inappropriate  if the state of the matter fields   were indeed $ \vert\zeta^{\textrm{(I)}}\rangle_{\textrm{target}}$, simply because  in  such case the  space-time  metric would  have to  be  different from the  one used  to make the  construction of that Hilbert space. 
In order to have a sensible picture, we need to connect this  state $\vert\zeta^{\textrm{(I)}}\rangle_{\textrm{target}}$ with another one 
 $\vert\zeta^{\textrm{(II)}}\rangle$ ``living" in a new Hilbert space $\mathscr{H}^{\textrm{(II)}}$ for which 
$\{g^{\textrm{(II)}},\hat{\varphi}^{\textrm{(II)}},\hat{\pi}^{\textrm{(II)}}, \mathscr{H}^{\textrm{(II)}},\vert\zeta^{\textrm{(II)}}\rangle\}$ is 
an actual SSC. We will denote the new SSC by SSC-II. 
Thus, first we need to determine the ``target'' (non-physical) state in $\mathscr{H}^{\textrm{(I)}}$ to which the initial state 
is in a sense ``tempted" to jump, and after that, we need to relate  such target state with a corresponding  state 
 in the Hilbert space of a new SSC, the SSC-II. We will define these  notions more precisely below.
Following our previous treatments on the subject 
(see for instance reference \cite{Perez:2005gh}), we will consider that the 
target state is chosen stochastically, guided by the quantum uncertainties of designated  field  operators,
evaluated on the initial state $\vert\xi^{\textrm{(I)}}\rangle$,  at the collapsing time. We  will  maintain the essence of such  prescriptions here.  
On the other hand,  regarding the identification between the two different SSCs involved in the collapse, 
there seems to be in principle  many natural options. The usefulness of those  depends, of course, on  the degree that  they actually determine 
 possible  SSC's.  In this paper we will be focusing  on the possibility  offered by  the 
following prescription: Consider that the collapse takes place along a Cauchy hypersurface $\Sigma_c$. 
A transition from the physical state $\vert\xi^{\textrm{(I)}}\rangle$ in $\mathscr{H}^{\textrm{(I)}}$ to the physical state 
$\vert\zeta^{\textrm{(II)}}\rangle$ in $\mathscr{H}^{\textrm{(II)}}$ (associated to the target \textit{non-physical} state 
$\vert\zeta^{\textrm{(I)}}\rangle_{\textrm{target}}$ in $\mathscr{H}^{\textrm{(I)}}$) will occur  in a way that
 \begin{equation}\label{recipe.collapses}
_\textrm{target}\langle\zeta^{\textrm{(I)}}\vert \hat{T}^{\textrm{(I)}}_{\mu\nu}[g^{\textrm{(I)}}, 
\hat{\varphi}^{\textrm{(I)}},\hat{\pi}^{\textrm{(I)}}]\vert\zeta^{\textrm{(I)}}\rangle_{\textrm{target}} \big|_{\Sigma_c}=
\langle\zeta^{\textrm{(II)}}\vert \hat{T}^{\textrm{(II)}}_{\mu\nu}[g^{\textrm{(II)}}, \hat{\varphi}^{\textrm{(II)}},\hat{\pi}^{\textrm{(II)}}]\vert\zeta^{\textrm{(II)}}\rangle \big|_{\Sigma_c}\, ,
\end{equation}
i.e.  in such a  way  that   the expectation value of the energy-momentum tensor associated to the states $\vert\zeta^{\textrm{(I)}}\rangle_{\textrm{target}}$ and 
$\vert\zeta^{\textrm{(II)}}\rangle$ evaluated on the Cauchy hypersurface $\Sigma_c$ coincide. Note that the 
 left hand side  in the expression  above 
is meant  to be constructed from the elements of the SSC-I (although $\vert\zeta^{\textrm{(I)}}\rangle_{\textrm{target}}$ is not really {\it the state} of the SSC-I), while 
the right hand side correspond to quantities  evaluated  using the SSC-II.
As we have relied for motivation of our proposal involving  the collapse  of the wave function on some ideas related  
to quantum gravity, and given that, at the classical level, the energy-momentum 
tensor acts as the ``source  of the gravitational interaction",
we find it reasonable to assume that it is precisely the expectation value of the energy-momentum tensor 
of the different states involved in the collapse the determining  aspect  for such identification.

That means that, in general, at the collapsing time we will have two different metrics $g^{\textrm{(I)}}$ and $g^{\textrm{(II)}}$ for a given Cauchy hypersurface. 
There could be some special situations for which both metrics $g^{\textrm{(I)}}$ and $g^{\textrm{(II)}}$ coincide over  some  
neighbourhood of  $\Sigma_c$, 
but
in general there is no reason to expect that there might  be a suitable interpolating metric description for the space-time
 during the  process of collapse. At best we might hope to find a recipe where  
 the  induced metric on the hypersurface would  be continuous at $\Sigma_c$, but  its normal  derivative (i.e. the extrinsic  curvature), would not.  
Later we will find that, indeed, this would be what happens in the case of interest when we use the matching prescription given in 
equation (\ref{recipe.collapses}).
By construction, the standard unitary 
evolution of quantum mechanics takes place within a given SSC, thus  we need the collapses in order to jump
between the different SSCs, and have the possibility of  describing  the  generation of   the  seeds of  cosmic  structure.
After all, generation means  that  ``something that did  not exist  at a certain  time does exist  at a later one''.
It  is  worth  emphasizing that the dynamical collapse of the wave function 
does not belong to what we could call ``well established
physics", although  there  are several  proposals  formulated  within the community working on foundational aspects
 of quantum theory (see for instance \cite{Bassi:2003gd} and references therein)  which  could
  be connected  with the general ideas outlined in this section.

We should note that the equation (\ref{EE}) would not, in general, 
hold through  the collapses.\footnote{In this  work we  are considering  
the collapse  as  taking place  instantaneously, i.e. on a  space like hypersurface
$\Sigma_c$. However, it is perhaps  better to think of this as an approximated  
description of something  taking place very fast in comparison to the other  
time  scales of the problem.}  
At such  times  the excitation of the  fundamental quantum gravitational degrees of freedom should be considered as  ``coming into play",
 with the corresponding breakdown of the semiclassical approximation. 
  In this context,  it is  worthwhile noting a certain  similitude  with the conclusions  of the analysis of the  resolution of the  
black hole singularity in loop quantum gravity \cite{Ashtekar}.  There, it  is  argued that   even though at the quantum level there is no singularity and  the region  has a perfectly appropriate  description in terms of loop  variables,  the  metric  description  simply does  not exist  for the region that would correspond to the singularity.
  Getting back to our case, this  possible breakdown 
 can be represented  at the formal  level by the inclusion of  a  term  $Q_{\mu\nu} $ in  
 the semiclassical field equations, which is supposed to become nonzero only during the  collapse
 of the quantum mechanical wave function of the matter fields:
\begin{equation}
\label{SemiCEQ}
  G_{\mu\nu} +Q_{\mu\nu}  =8\pi G  \langle \hat{T}_{\mu\nu}  \rangle .
  \end{equation}

  The  setting is  so far general  and would allow in principle to consider  various situations. 
In this work, we  want to focus on the  description  of the emergence of the  seeds of cosmic structure in  the context  of a universe 
which was initially  described by a homogeneous and isotropic state for the gravitational and matter degrees of freedom.   
 The idea  is  that, at some point, the quantum state of the matter fields reaches
a stage whereby the corresponding state for the gravitational degrees of freedom leads to  a quantum jump of the matter field wave function.
The resulting state of the matter fields needs not  share  the symmetries of the initial state, 
\begin{equation}
\vert\textrm{symmetric}\rangle \to \vert\textrm{non-symmetric}\rangle,
\end{equation}
and its connection to the  gravitational degrees of freedom,  which again is assumed to be 
accurately described by the Einstein semiclassical  equations,  leads to a 
geometry that is no longer homogeneous and isotropic.
The matching at the collapsing time of the two SSCs constructed in 
Sections \ref{before the collapse} and \ref{After the collapse} will be explicitly analysed in Section \ref{collapse}
following the general ideas discussed here.

 
  
\section{Fitting inflation into our general formal  scheme}\label{formalizing}

In this section   we  will show  a detailed realization of the ideas  
described  above. The  motivation for this  is  twofold: On the one  hand  it  
will serve  as  a  proof  of concept,  and  on the other  it will help  us  
clarify the  description of the
cosmological  situation  that gave rise to the  general ideas  explored  in \cite{Perez:2005gh} and  subsequent  works.

As  we have argued, in order to have a sensible picture for the problem at hand 
we should determine the SSC's suitable for describing the cosmological evolution of the universe. 
Our point of view is that,  in principle, something  like this could be done in most  situations. 
However, it seems clear  that, in practice,  the complexity of   
the problems one is  commonly interested in would  make  this task simply unmanageable.
Nevertheless, as we are mainly interested in the study of the relatively simple case of  the very early universe, we can restrict ourselves
to the subset of the {\it nearly} (as characterized  by the  parameter $\varepsilon$ in expression (\ref{fourier.psi})) homogeneous and isotropic SSCs.
%
That is,  even though the  problem  is   well defined in general, we   will approach its solution  
through a practical perturbative approach.  
However, we will always 
have under control the exact nature of the approximations we will be using, 
providing for the first time
a  clear interpretative picture which is well defined from the beginning and 
does not change as  the discussion progresses.

We  will start now to  be rather  specific. 
As it was indicated in the previous section the matter fields will  be treated  in the  language of a 
quantum field theory  on curved  space-times \cite{Birrell1984, Wald1994, Mukhanov2007, Parker2009}. 
We  will concentrate here on the case of  a single  scalar field, the inflaton. 
The other fields, and in particular all the fields of  the standard model of particle 
physics, are assumed in their vacuum state and will  be ignored in the present work.
In order to  clarify  our  ideas, let  us remind the reader that inflation is  supposed to  occur in a patch   which  emerges  from the Planck  era 
(at time $ t_{\textrm{IS}}$, where inflation starts),
  corresponding to a situation  where the scalar field is in a regime  where the  inflaton potential  is sufficiently  large,  
the kinetic term sufficiently small, and the space-time geometry
sufficiently close to a homogeneous and isotropic one for inflation dynamics to take place. 
 After the onset of such dynamics, that patch exponentially inflates, leading to  a 
 region of space-time which is very close to a homogeneous  and isotropic 
  flat Robertson-Walker universe, with the remnants of the inhomogeneities that 
were present at $t_{\textrm{IS}}$ reduced by an exponential factor in the number of e-folds, 
$\mathcal{N}\equiv \ln (a/a_{\textrm{IS}})$, with $a$ the value for the scale factor (see expression (\ref{FRW perturbed}) below). 
Thus, there  would  be at any time during inflation 
 remnants of   inhomogeneity that are at most of  order  $ e^{-\mathcal{N}}$. 
These remanent  inhomogeneities    are not supposed
   to  be relevant at all in  any physical  consideration  during  much of the inflationary  regime  itself,
and are thus  said to  be  essentially   erased by  inflation.
   This  is the situation  where   our analysis  will be focussed on, and
    this is indeed necessary if we want to consider as  justified  any deduction that 
relies on the  use of the vacuum  state of the inflaton field    to  characterize the seeds  of 
 the cosmological structures we observe today. Alternatively, if  one  wanted to 
claim that such  remnants  from the pre-inflationary era are tied to the   
generation of the  seeds  of  structure,  we  would have to accept that it is 
impossible to  predict  any features  of  such structure, because we do not know  
anything  about  the form of the  spectrum  that  might characterize  such  
remnants.

The fact  that we do not even have a truly non-perturbative  method  for dealing  with interacting quantum  fields in curved spaces 
leads  us to set our attention on the case of a massive, non-interactive scalar field.
  In fact, such a model applied to the  early phase of accelerated expansion has been extensively analysed in the
 literature, see for instance \cite{Belinsky:1985zd}. At the classical level the inflaton field 
satisfies the Klein-Gordon equation,
\begin{equation}\label{eq.dynamical}
g^{\mu\nu}\nabla_{\mu}\nabla_{\nu}\phi-m^2\phi=0,
\end{equation}
with 
$g^{\mu\nu}$ the inverse of the space-time metric,
and $\nabla_{\mu}$ the covariant derivative.
Given  $\phi_1(x)$ and $\phi_2(x)$ two solutions to the classical equation of motion (\ref{eq.dynamical}), the symplectic product 
is defined by 
\begin{equation}\label{scalar.product}
(\phi_1,\phi_2)_{\textrm{Sympl}}\equiv-i\int_{\Sigma}\left[\phi_1(\partial_\mu\phi^*_2)-(\partial_\mu\phi_1)\phi^*_2\right]\,d\Sigma^\mu.
\end{equation}
Here $d\Sigma^\mu\equiv n^\mu d\Sigma$, with $n^\mu$ a time-like, future-directed, normalized 4-vector orthogonal to the 3-dimensional 
Cauchy hypersurface $\Sigma$, and $d\Sigma =\sqrt{g_{\Sigma}}d^3x$ its volume element.
As usual the expression (\ref{scalar.product}) does not depend on the selected 
Cauchy hypersurface. 
In terms of the
conjugate momentum associated to the inflaton field, $\pi(x)\equiv\sqrt{g_{\Sigma}}\left(n^\mu\partial_\mu\phi\right)$, the symplectic  product  
can be re-written in the form
\begin{equation}\label{scalar.product.pi}
((\phi_1,\pi_1),(\phi_2,\pi_2))_{\textrm{Sympl}} \equiv-i\int_{\Sigma}\left[\phi_1\pi^*_2-\pi_1\phi^*_2\right]\,d^3 x.
\end{equation}

At the quantum level the inflaton field and its conjugate momentum are  promoted to field operators acting on a 
Hilbert space $\mathscr{H}$. These operators  must  satisfy  the standard equal time 
commutation relations between them, which,  upon   an appropriate   choice of   space-time coordinates (to be further specified  shortly),  take the form,
%
\begin{equation}\label{standard.comm}
[\hat{\phi}(\eta,\vec{x}),\hat{\pi}(\eta,\vec{y})]=i\hbar \delta(\vec{x}-\vec{y}),\quad [\hat{\phi}(\eta,\vec{x}),\hat{\phi}(\eta,\vec{y})]=
[\hat{\pi}(\eta,\vec{x}),\hat{\pi}(\eta,\vec{y})]=0.
\end{equation}
The standard way to proceed now is to decompose 
$\hat{\phi}(x)$ in terms of the time-independent creation 
 and annihilation 
  operators,
\begin{equation}\label{exp}
 \hat{\phi}(x)=\sum_\alpha \left(\hat{a}_{\alpha}u_{\alpha}(x)+\hat{a}^{\dagger}_{\alpha} u^*_{\alpha}(x)\right),
\end{equation}
with the functions $u_{\alpha}(x)$ a  complete set of normal modes orthonormal with respect to the symplectic product, 
i.e.
\begin{equation}\label{dyn.prod}
(g^{\mu\nu}\nabla_{\mu}\nabla_{\nu}-m^2)u_{\alpha}=0,\quad
(u_{\alpha},u_{\alpha'})_{\textrm{Sympl}}=\hbar \delta_{\alpha\alpha'}.
\end{equation}
A similar expression to that given in (\ref{exp}) can be obtained for the conjugate momentum $\hat{\pi}(x)$,
 replacing the functions $u_{\alpha}$ by $\sqrt{g_{\Sigma}} (n^\mu\partial_\mu u_{\alpha} )$.
As  we will be only interested in  configurations  corresponding to space-times very close to a  flat Robertson-Walker one,
  we will be  able  to  choose the  labels $\alpha$ specifying the different mode solutions
to  be  the  wave   vectors  $\vec{k}$,  despite the fact that  the spatial sections are not, in general, exactly  
flat. 
Clearly, one should always keep in mind that, 
in general, the functions $u_{\vec{k}}(x)$ would  not correspond
  exactly to the standard Fourier modes in flat space
(the  simple functional form of the modes $u_{\vec{k}}(x)\propto e^{i \vec{k}\cdot\vec{x}}$ will be appropriate  only for the exactly  homogeneous
 and isotropic case).

With all these conventions (expressions (\ref{exp}) and (\ref{dyn.prod}) above)
the commutators (\ref{standard.comm}) translate into the standard
\begin{equation}
 [\hat{a}_{\vec{k}},\hat{a}^{\dagger}_{\vec{k}'}]=
 \delta_{\vec{k}\vec{k}'},\quad [\hat{a}_{\vec{k}},\hat{a}_{\vec{k}'}]=[\hat{a}^{\dagger}_{\vec{k}},\hat{a}^{\dagger}_{\vec{k}'}]=0
\end{equation}
for the creation and annihilation operators.
As usual, the vacuum is defined to  be the state that is annihilated by 
all the $\hat{a}_{\vec{k}}$'s, (i.e. $\hat{a}_{\vec{k}}\vert 0\rangle=0$ for all $\vec{k}$).
The Hilbert space can be constructed ({\it \`a la} Fock) by successive applications of creation operators on the vacuum.
However, the relations (\ref{dyn.prod}) do not determine the set of mode solutions unequivocally, and the particular 
choice corresponds to an election of the vacuum.

This construction for the Hilbert space of the inflaton field is generic and applies for any (globally hyperbolic) space-time configuration.
An exact solution will not be possible in general, but an approximate one suitable
 for the study of the very early universe could be obtained perturbatively.
The study  of  the seeds of cosmic structure depends essentially on the scalar sector of the perturbations.  
Ignoring for simplicity 
the so called vector and tensor modes,  
we will choose a coordinate system (the so called conformal Newtonian gauge) in which
the space-time metric takes the form 
\begin{equation}
ds^{2}=a^{2}(\eta)\left[-(1+2\psi)d\eta^{2}+(1-2\psi)\delta_{ij}dx^{i}dx^{j}\right],\quad\textrm{with}\quad\psi(\eta,\vec{x})\ll 1\label{FRW perturbed}.
\end{equation}
Here $a$ is the scale factor, $\psi$  an analogue to the 
Newtonian potential, and $\eta$ the cosmological time in conformal coordinates.
The coordinates $x^i$ label the observers co-moving with the expansion. 
 Note that, in accordance  with  our approach, both $a(\eta)$ and $\psi(\eta,\vec{x})$ are (dimensionless) {\it classical} fields. 
 Setting  $\psi=0$ we recover a spatially flat homogeneous and isotropic 
 Robertson-Walker universe.\footnote{Indeed, we recover that situation even if take
$\psi=\psi(\eta)$, as can be seen by a simple change of the space-time coordinates. 
Here we will impose $\int d^3\vec{x}\,\psi(\eta,\vec{x})=0$ in order to remove  those ambiguities.}
 In order to avoid problems in the infrared we will  work with periodic boundary conditions over a 
 box of size $L$. We can take the limit $L\to \infty$ at the end of the calculations. 
In the present context   we   can describe the Newtonian potential  in terms of  a Fourier decomposition  and write  generically
\begin{equation}\label{fourier.psi}
\psi(\eta,\vec{x})= \varepsilon \sum_{\vec{k}\neq 0} 
\tilde{\psi}_{\vec{k}}(\eta)e^{i\vec{k}\cdot\vec{x}},
\end{equation}
where  the sum is  over all vectors $\vec{k}$
 with $k_n=2\pi j_n/L$, $j_n=0,\pm 1,\pm 2,\ldots$ and $n=1,2,3$. 
In order to have a real value for $\psi(\eta,\vec{x})$ we should demand $\tilde{\psi}_{\vec{k}}(\eta)=\tilde{\psi}_{-\vec{k}}^*(\eta)$, and 
we will be assuming $\tilde{\psi}_{\vec{k}}(\eta) \lesssim \mathcal{O}(1)$ and $\varepsilon\ll 1$, so we can 
guarantee $\psi(\eta,\vec{x})\ll 1$.
Note that as the space-independent part of the Newtonian potential can be reabsorbed in the scale factor, it does not appear in (\ref{fourier.psi}).
Recall also  that we were working in co-moving coordinates, where the $\vec{k}$'s are fixed in time and label each particular mode 
(related to their physical values through $\vec{k}/a$).

Working  up to the first order in $\varepsilon$, 
 the equations (\ref{dyn.prod}) 
simplify to  
\begin{subequations}\label{dyn.ort.linear}
\begin{equation}
(1-2\psi)(\ddot{u}_{\vec{k}}+2\mathcal{H}\dot{u}_{\vec{k}})-(1+2\psi)\Delta u_{\vec{k}}-4\dot{\psi}\dot u_{\vec{k}}+a^2m^2 u_{\vec{k}} =0,
\end{equation}
\begin{equation}\label{symp.linear}
\int_{\eta=\textrm{const.}}\left[u_{\vec{k}}(\partial_{\eta}u_{\vec{k}'}^*)-(\partial_{\eta}u_{\vec{k}})u_{\vec{k}'}^*\right](1-4\psi)d^3x = i\hbar a^{-2}\delta_{\vec{k}\vec{k}'}.
\end{equation}
\end{subequations}
For the general case even the construction of that SSC will not be  trivial, although, in principle, it  would  
be doable. However,  in order to proceed,  we will   consider  a simple example: the  transition from   
an exactly  homogeneous and isotropic universe, 
$\psi(\eta, x)=0$,
to a  situation where,  as a result of one  of our  collapse  events, a single nontrivial  plane-wave  is excited. We characterize that  by the wave  vector $\vec{k}_0$,  and write  
$\psi(\eta, x) = \varepsilon\tilde{\psi}_{\vec{k}_0}(\eta) e^{i\vec{k}_0\cdot\vec{x}}+c.c.$, with  $\varepsilon \ll 1$ and $c.c.$ denoting the complex conjugate. 
Next, we  proceed to explicitly construct the  two SSC's: the first   corresponding to the  homogeneous and isotropic pre-collapse  
situation,  and the second to the post-collapse one, characterized  by a SSC  where there is an actual fluctuation  with  wave vector $\vec{k}_0$. 
Then,  we  will consider in some detail    the   character of 
 the  description  of the transition between those  two. 
 
A small comment on notation is in order here. The objects that are specific to  
the constructions we will be  discussing in Sections \ref{before the collapse}
and \ref{After the collapse} will bear an index (I or II) indicating  
the specific SSC construction to which they belong. 
In reading those two sections the reader can simply ignore that index,  
as they  will not change within each section. However,
we have left the index in place in order to avoid confusion when the two  
constructions are brought together in the discussion of their matching in
Section \ref{collapse}.

\subsection{A homogeneous and isotropic SSC}\label{before the collapse}

At a time corresponding to few  e-foldings after inflation started the  relevant region  of the  universe is thought to be described by
 a homogeneous and isotropic SSC. The Newtonian potential vanishes at that time, $\psi(\eta, \vec{x})=0$. 
We  will call it the  first SSC, or simply the SSC-I. 
 In order to carry out that construction we will take the space-time metric 
to  be that of a Robertson-Walker universe, with a pre-established (nearly) 
de Sitter scale factor. 
The small deviation from the exact de Sitter expansion will be 
parametrized by 
$\epsilon^{\textrm{(I)}}\equiv 1-\dot{\mathcal{H}}^{\textrm{(I)}}/\mathcal{H}^{2\textrm{(I)}}$, 
where $\mathcal{H}^{\textrm{(I)}}\equiv\dot a^{\textrm{(I)}}/a^{\textrm{(I)}}$ is a 
measure of the expansion rate of the universe, related to the standard 
Hubble parameter by $H^{\textrm{(I)}}=\mathcal{H}^{\textrm{(I)}}/a^{\textrm{(I)}}$.
During slow-roll $0<\epsilon^{\textrm{(I)}}\ll 1$
(not to be confused with $\varepsilon\ll1$).
For practical purposes and for all the situations analysed in this paper
it will be sufficient to consider the scale factor
$a^{\textrm{(I)}}(\eta) = (-1/H^{\textrm{(I)}}_0\eta)^{1+\epsilon^{\textrm{(I)}}}$
and
the Hubble parameter in conformal coordinates 
$\mathcal{H}^{\textrm{(I)}}=-(1+\epsilon^{\textrm{(I)}})/\eta$
to the first order in the slow-roll parameter (remember that $-\infty <\eta < 0$).
Next we  need to  find a complete  set 
of normal modes $u^{\textrm{(I)}}_{\vec{k}}(x)$ 
for the previously given homogeneous and isotropic space-time configuration.
Naturally, given the   symmetries of the spatial background, we can look for  solutions  of the form
\begin{equation}\label{Fourier.dec}
 u^{\textrm{(I)}}_{\vec{k}}(x)=v^{\textrm{(I)}}_{\vec{k}}(\eta) e^{i\vec{k}\cdot\vec{x}}/L^{3/2}.
\end{equation}
Introducing $\psi=0$ and the ansatz (\ref{Fourier.dec}) into the equations (\ref{dyn.ort.linear}) we obtain
\begin{subequations}\label{dyn.orth.HI}
\begin{equation}
\ddot{v}^{\textrm{(I)}}_{\vec{k}}+2\mathcal{H}^{\textrm{(I)}}\dot{v}^{\textrm{(I)}}_{\vec{k}}+\left(k^2+a^{2\textrm{(I)}}m^2\right)v^{\textrm{(I)}}_{\vec{k}}=0,\label{dynamical.zeroth}
\end{equation}
\begin{equation}
 v^{\textrm{(I)}}_{\vec{k}} { \dot{v}^{\textrm{(I)}*}_{\vec{k}} }-\dot{v}^{\textrm{(I)}}_{\vec{k}}{v^{\textrm{(I)}*}_{\vec{k}} }=i\hbar a^{-2\textrm{(I)}}.\label{orthonormal}
\end{equation}
\end{subequations}
We will find that $a^{2\textrm{(I)}}m^2$ is first order in the slow-roll parameter,
$a^{2\textrm{(I)}}m^2=3\epsilon^{\textrm{(I)}}\mathcal{H}^{2\textrm{(I)}}$
to the first nonvanishing order in $\epsilon^{\textrm{(I)}}$ (see equation (\ref{epsilon1.m}) bellow),
so equation (\ref{dynamical.zeroth}) should be considered to this same order, i.e. we should take $\mathcal{H}^{\textrm{(I)}}=-(1+\epsilon^{\textrm{(I)}})/\eta$.
For the modes with $k\neq 0$ the most general 
solution to the equation (\ref{dynamical.zeroth}) is a linear combination of the functions
$\eta^{3/2+\epsilon^{\textrm{(I)}}}H_\nu^{(1)}(-k\eta)$ and $\eta^{3/2+\epsilon^{\textrm{(I)}}}H_\nu^{(2)}(-k\eta)$, with  $H_\nu^{(1)}(-k\eta)$ and $H_\nu^{(2)}(-k\eta)$
the Hankel functions of first and second kind, and 
$\nu^{\textrm{(I)}}=3/2+\epsilon^{\textrm{(I)}}-m^2/3H^{2\textrm{(I)}}_0$
to the first order in the slow-roll parameter.
A choice of modes corresponds to  an election of the vacuum. For a space-time background without a time-like killing
 vector field there is no preferential choice, but following the standard literature on the subject we  will  take the Bunch-Davis 
convention:  i.e.  use  modes  such  that in the asymptotic past they behave as  purely  ``positive frequency solutions",   
 normalized according to (\ref{orthonormal}). 
Working to the lowest non-vanishing order in the slow-roll parameters ($\nu^{\textrm{(I)}}=3/2$), 
we obtain
\begin{equation}\label{standard}
  v_{\vec{k}}^{\textrm{(I)}}(\eta)=\sqrt{\frac{\hbar}{2k}}\left(-H^{\textrm{(I)}}_0\eta\right)\left(1-\frac{i}{k\eta} \right)e^{-ik\eta}.
\end{equation}
These functions coincide with the standard mode solutions of a massless ($m=0$) scalar field in a de Sitter ($\epsilon^{\textrm{(I)}}=0$) universe.
%
%
We note,  however, the Hankel functions are not well behaved at the origin, and thus  the zero mode is not included in (\ref{standard}). 
For  $k=0$ the general solution to the equation (\ref{dynamical.zeroth}) is a 
linear combination of the functions $\eta^{(3+2\epsilon^{\textrm{(I)}}-2\nu)/2}$ 
and $\eta^{(3+2\epsilon^{\textrm{(I)}}+2\nu)/2}$. 
The choice is  arbitrary, provided it  has  positive symplectic norm. Here we  take 
\begin{equation}\label{u.zero.lowest}
 v_0^{\textrm{(I)}}(\eta)=\sqrt{\frac{\hbar}{H^{\textrm{(I)}}_0}}\left[1-\frac{i}{6}
\left(-H^{\textrm{(I)}}_0\eta\right)^{3}\right]\left(-H^{\textrm{(I)}}_0\eta\right)^{m^2/3H^{2\textrm{(I)}}_0},
\end{equation}
%
which  has been normalized using the condition (\ref{orthonormal}).
Note that, contrary to what we did for the $k\neq 0$ modes, 
we have worked now to the first order in the slow-roll parameter.
We will find that this is necessary in order to accommodate a slow-rolling 
expectation value for the zero mode.

So far we have 
given a prescription to construct the  mode solutions 
$u_{\vec{k}}^{\textrm{(I)}}(x)$ for the SSC-I, and thus we have a construction of 
the Hilbert space and a representation of the fundamental fields as operators  
acting  on it. 
However, we still need to find a state $\vert\xi^{\textrm{(I)}}\rangle$ in $\mathscr{H}^{\textrm{(I)}}$ such that its expectation 
value for the energy-momentum tensor leads to the desired nearly de Sitter, homogeneous and isotropic cosmological expansion.
As expected, 
the state in the SSC-I can just have the zero mode excited. 
The expectation value of the field operator is then homogeneous and isotropic,
\begin{equation}\label{SV.SSCI}
\phi^{\textrm{(I)}}_{\xi,0}(\eta)\equiv
\langle\xi^{\textrm{(I)}}\vert\hat{\phi}^{\textrm{(I)}}(x)\vert\xi^{\textrm{(I)}}\rangle=\xi^{\textrm{(I)}}_0 v_0^{\textrm{(I)}}(\eta) /L^{3/2}+c.c.,
\end{equation}
with $\xi^{\textrm{(I)}}_0\equiv\langle\hat{a}_0^{\textrm{(I)}}\rangle$. The subindex ``0" in $\phi^{\textrm{(I)}}_{\xi,0}(\eta)$ makes reference 
to this fact,
whereas the super-index ``(I)" indicates that the  field operators involved are those of the SSC-I, and  also that the expectation
  values  are  taken in the state $\vert\xi^{\textrm{(I)}}\rangle$.
We should  note that as the potential is quadratic in the field operator the Ehrenfest theorem 
guarantees the classical equation of motion for the expectation value of the inflaton field.

As it was just argued, the symmetries of the space-time background lead us to consider  a  state in which all the modes 
with $k\neq 0$ are in their vacuum state, while the zero mode is  excited. Thus, we  consider a state of the form
\begin{equation}\label{state.before}
\vert\xi^{\textrm{(I)}}\rangle= {\cal F}(
\hat{a}^{\textrm{(I)}\dagger}_0) \vert 0 
^{\textrm{(I)}}\rangle,
\end{equation}
where $ {\cal F}(\hat{X})$  stands  for a suitable generic  function acting on  the operators $\hat X$ 
(as  we will see it  will  be sufficient for our purposes in this paper to  use the   function that  is associated  with the ``coherent'' states, namely  $ {\cal F}(\hat{X})  \propto\textrm{exp}(\hat{X})$).  
For those particular states with only the zero mode excited, the components  $00$ and $i=j$  of Einstein's field equations simplify to\footnote{There is also an infinite contribution to the 
expectation value of the energy-momentum tensor coming from the vacuum of the theory,  but this is  an issue related to the well known 
cosmological constant problem and it will not be considered any further here. We could assume that the energy-momentum tensor is 
renormalized,  say  {\it \`a la} Haddamard, and that the  cosmological constant is  set to  zero. 
In practice we will simply impose normal ordering on the energy-momentum tensor operator,
$:\hat{a}_{\vec{k}}\hat{a}_{\vec{k}}^{\dagger}:=\hat{a}_{\vec{k}}^{\dagger}\hat{a}_{\vec{k}}$.}
\begin{subequations}\label{einstein.HI}
 \begin{eqnarray}
  3\mathcal{H}^{2\textrm{(I)}} &=& 4\pi G\left((\dot{\phi}^{2\textrm{(I)}})_{\xi,0}+a^{\textrm{2(I)}}m^2(\phi^{2\textrm{(I)}})_{\xi,0} \right),\label{einstein.HI.1}\\
  \mathcal{H}^{2\textrm{(I)}}+2\dot{\mathcal{H}}^{\textrm{(I)}}  &=& -4\pi G\left((\dot{\phi}^{2\textrm{(I)}})_{\xi,0}-a^{2\textrm{(I)}}m^2(\phi^{2\textrm{(I)}})_{\xi,0} \right).\label{einstein.HI.2}
 \end{eqnarray}
\end{subequations}
The  two sides of  the $0i$ and the $i\neq j$ equations vanish identically for that particular metric  and  state.
Here $(\dot{\phi}^{2\textrm{(I)}})_{\xi,0}\equiv\langle\xi^{\textrm{(I)}}\vert:(\partial_\eta\hat{\phi}^{\textrm{(I)}})^2:\vert\xi^{\textrm{(I)}}\rangle$ and 
$(\hat{\phi}^{2\textrm{(I)}})_{\xi,0}\equiv\langle\xi^{\textrm{(I)}}\vert:(\phi^{\textrm{(I)}})^2:\vert\xi^{\textrm{(I)}}\rangle$ are functions of $\eta$. 
The equations (\ref{einstein.HI}) are analogous (but \textit{not}  exactly equal) to those obtained in the context of a classical field theory, 
with the squares of the scalar field and its time derivative replaced now by the expectation value of their corresponding  operators. 
This difference could, in principle,  introduce important departures from the classical behaviour, but fortunately this problem will not affect us in the simple case  treated here.
As we have just mentioned,  Ehrenfest's theorem guarantees the classical equations 
of motion for the expectation value of the inflaton field.
In general, the 
 ``classical relations"  
$(\dot{\phi}^{2\textrm{(I)}})_{\xi,0} =(\dot{\phi}^{\textrm{(I)}}_{\xi,0})^2$ and $(\phi^{2\textrm{(I)}})_{\xi,0} =(\phi^{\textrm{(I)}}_{\xi,0})^2$
will not  hold 
(that is, $\langle\hat{a}_0^{\textrm{(I)}}\rangle=\xi^{\textrm{(I)}}_0$ does not necessarily imply $\langle\hat{a}_0^{2\textrm{(I)}}\rangle=\xi^{2\textrm{(I)}}_0$).
However, as we
want to consider a state $\vert\xi^{\textrm{(I)}}\rangle$ that is sharply peeked around a classical field configuration, 
$\phi_{\xi,0}^{\textrm{(I)}}(\eta)$ and $\pi_{\xi,0}^{\textrm{(I)}}(\eta)$, we take 
it to correspond to a ``highly excited coherent" state (i.e. $\hat{a}^{\textrm{(I)}}_0\vert\xi^{\textrm{(I)}}\rangle=\xi^{\textrm{(I)}}_0\vert\xi^{\textrm{(I)}}\rangle$, with $\xi^{\textrm{(I)}}_0\in\mathbb{C}$),
for which those  classical relations do hold.
In that case we recover precisely the same  Friedmann equations of the standard treatments.

Requiring that the universe is characterized by a regime of slow-roll inflation 
with expansion rate $H_0^{\textrm{(I)}}$ and slow-roll parameter $\epsilon^{\textrm{(I)}} \equiv 1-\dot{\mathcal{H}}^{\textrm{(I)}}/\mathcal{H}^{2\textrm{(I)}}$
implies $3(\dot{\phi}_{\xi,0}^{\textrm{(I)}})^2 =\epsilon^{\textrm{(I)}} a^{2\textrm{(I)}}m^2(\phi_{\xi,0}^{\textrm{(I)}})^2$.
This expression is obtained, working up to the first order in the slow-roll parameter $\epsilon^{\textrm{(I)}}$, by combining the 
equations (\ref{einstein.HI.1}) and (\ref{einstein.HI.2}) (once we have assumed a coherent state).
  Solving this equation for $\phi_{\xi,0}^{\textrm{(I)}}(\eta)$ we obtain
\begin{equation}\label{u0.s.r.}
 \phi_{\xi,0}^{\textrm{(I)}}(\eta)\propto\eta^{\sqrt{\epsilon^{\textrm{(I)}} m^2/3H^{\textrm{2(I)}}_0}}.
\end{equation}
  Comparing  with  expressions (\ref{u.zero.lowest}) and (\ref{SV.SSCI}),  and  taking   the  parameter $\xi_0^{\textrm{(I)}}$
  as real, 
   \begin{equation}\label{exp.phi}
\langle\xi^{\textrm{(I)}}\vert\hat{\phi}^{\textrm{(I)}}(x)\vert\xi^{\textrm{(I)}}\rangle=\frac{2\xi^{\textrm{(I)}}_0}{L^{3/2}}
\sqrt{\frac{\hbar}{H^{\textrm{(I)}}_0}}
\left(-H^{\textrm{(I)}}_0\eta\right)^{m^2/3H^{2\textrm{(I)}}_0}.
\end{equation}
Thus, we  find compatibility of equations 
 (\ref{u0.s.r.}) and (\ref{exp.phi}), corresponding to a period of slow-roll inflation with 
\begin{equation}\label{epsilon1.m}
\epsilon^{\textrm{(I)}}=\frac{m^2}{3H^{2\textrm{(I)}}_0}, \quad H^{\textrm{(I)}}_{0} =2\epsilon^{\textrm{(I)}} t_p^2 \frac{(\xi^{\textrm{(I)}}_0)^2}{L^3},
\end{equation}
where  $t_p^2 = 8\pi G\hbar$ stands for  the Planck time.\footnote{The presence in 
the last  expression  of the factor $L^{-3} $ might seem  strange at first sight,  
as  the  size  of the  artificial 
 box should have no impact on the expansion rate of the universe. However, we  must recall that  in our  notation  the  expectation value of the scalar field is 
 proportional to  $L^{-3/2}$,  as can be  seen in  expression (\ref{SV.SSCI}). 
Thus, what  we have here  is  simply   the fact that 
$H^{\textrm{(I)}}_{0}$  is proportional to  $\langle \hat{\phi}^2 \rangle$.}
As it is  well known, in order to have $\epsilon^{\textrm{(I)}}\ll1$ we 
must require $m\ll H_0^{\textrm{(I)}}$. 
Inflation is expected to take place at very high energies, so this requirement is not  generally taken as problematic.
On the other hand, given the values of the parameters $H_0^{\textrm{(I)}}$ and 
$\epsilon^{\textrm{(I)}}$ characterizing the
cosmological expansion during inflation, we can read from the second expression in 
(\ref{epsilon1.m}) the state for the zero mode, $\xi^{\textrm{(I)}}_0$ 
(or more precisely, $\xi^{\textrm{(I)}}_0/L^{3/2}$). This fixes the state for the 
SSC-I, as all  other modes are taken to be in  their vacuum state (with respect 
to the Bunch-Davies convention). We thus have the metric, the quantum field theory 
construction, and the specific state that is compatible with Einstein's equation.
 This therefore completes the  construction of the SSC-I.
 We now turn to the  more  complex case corresponding to a  slightly inhomogeneous  and  anisotropic situation.
 
\subsection{A SSC with the $\vec{k}_0$-mode excited}\label{After the collapse}

Here  we want to carry out the construction of a new SSC corresponding to an excitation in the Newtonian potential characterized by the wave-vector $\vec{k}_0$.  We  will denote this new SSC by SSC-II.
We must  consider a  slight  deviation from the previous homogeneous  and isotropic cosmological background, 
characterized now by the parameters $H_0^{\textrm{(II)}}$ and $\epsilon^{\textrm{(II)}}$ (which  might, in principle,  differ slightly 
from those corresponding to the SSC-I  discussed  in the previous   section), 
 and a Newtonian potential with  one  single term excited in the expression (\ref{fourier.psi}), described by  an 
(in  principle)  arbitrary function 
$\tilde \psi_{\vec{k}_0}(\eta) = P(\eta)$.
We will latter see that the semiclassical field equations will allow us to 
obtain this function and the quantum state of the SSC, leading to a complete  
determination of this construction.

However, the first step in the building of   the SSC-II will be  the construction  of  the quantum theory for the inflaton field on   
the   new type  of  space-time configuration we are considering now (with any  given $P(\eta)$).  
That is,  we need a  complete set of normal modes $u_{\vec{k}}^{\textrm{(II)}}(x)$ appropriate for the new construction.
As  we  are working up  to the first order in $\varepsilon$,  and since the  Newtonian potential is given by
$\psi(\eta,\vec{x})=
\varepsilon P(\eta)e^{i\vec{k}_0\cdot{ \vec x}}+c.c. $,  
we can consider the ansatz:
\begin{equation}\label{ansatz}
 u_{\vec{k}}^{\textrm{(II)}}(x)= \varepsilon\delta v_{\vec{k}}^{\textrm{(II)}-}(\eta)\,e^{i(\vec{k}-\vec{k}_0)\cdot\vec{x}}/L^{3/2}+v^{\textrm{(II)0}}_{\vec{k}}(\eta)\, e^{i\vec{k}\cdot\vec{x}}/L^{3/2}
+\varepsilon\delta v_{\vec{k}}^{\textrm{(II)}+}(\eta)\,e^{i(\vec{k}+\vec{k}_0)\cdot \vec{x}}/L^{3/2}.
\end{equation}
Introducing (\ref{ansatz}) into (\ref{dyn.ort.linear})  we find that,  to the zeroth order in the $\varepsilon$, the   
evolution equation  is given by
\begin{subequations}
\begin{equation}
\ddot{v}^{\textrm{(II)}0}_{\vec{k}}+2\mathcal{H}^{\textrm{(II)}}\dot{v}^{\textrm{(II)}0}_{\vec{k}}+\left(k^2+a^{2\textrm{(II)}}m^2\right)v^{\textrm{(II)}0}_{\vec{k}}=0,\label{hom.SSCII}
\end{equation}
 with normalization condition
\begin{equation}
 v^{\textrm{(II)}0}_{\vec{k}} { \dot{v}^{\textrm{(II)}0*}_{\vec{k}} }-\dot{v}^{\textrm{(II)}0}_{\vec{k}}{ v^{\textrm{(II)}0*}_{\vec{k}} }=i\hbar a^{-2\textrm{(I)}},\label{normalization.SSCII.0}
\end{equation}
\end{subequations}
while at first order in $\varepsilon$ the  corresponding evolution equation takes the form
\begin{subequations}
\begin{equation}
\delta\ddot{v}^{\textrm{(II)}\pm}_{\vec{k}}+2\mathcal{H}^{\textrm{(II)}}\delta\dot{v}^{\textrm{(II)}\pm}_{\vec{k}}+\left[(\vec{k}\pm\vec{k}_0)^2+a^{2\textrm{(II)}}m^2\right]\delta v^{\textrm{(II)}\pm}_{\vec{k}}=
F^{\pm}_{\vec{k}}(\eta),\label{eq.deltau.1}
\end{equation}
where, 
\begin{eqnarray}
F^{+}_{\vec{k}}(\eta)&\equiv& 4\dot{P} \dot{v}^{\textrm{(II)0}}_{\vec{k}}-2\left(2k^2+a^\textrm{2(II)}m^2\right) P v^{\textrm{(II)0}}_{\vec{k}},\\
F^{-}_{\vec{k}}(\eta)&\equiv& 4\dot{P^*} \dot{v}^{\textrm{(II)0}}_{\vec{k}}-2\left(2k^2+a^\textrm{2(II)}m^2\right)P^{*} v^{\textrm{(II)0}}_{\vec{k}}.
\end{eqnarray}
The  normalization  condition is given by
\begin{equation}\label{orth.first}
\dot{v}_{\vec{k}+\vec{k}_0}^{\textrm{(II)0}*}\delta v_{\vec{k}}^{\textrm{(II)}+}-v_{\vec{k}+\vec{k}_0}^{\textrm{(II)0}*}\delta\dot{v}_{\vec{k}}^{\textrm{(II)}+} 
-\dot{v}_{\vec{k}}^{\textrm{(II)0}}\delta v_{\vec{k}+\vec{k}_0}^{\textrm{(II)}-*}+v_{\vec{k}}^{\textrm{(II)0}}\delta\dot{v}_{\vec{k}+\vec{k}_0}^{\textrm{(II)}-*}
=  4\left(v_{\vec{k}}^{\textrm{(II)0}}\dot{v}_{\vec{k}+\vec{k}_0}^{\textrm{(II)0}*}-\dot{v}_{\vec{k}}^{\textrm{(II)0}}v_{\vec{k}+\vec{k}_0}^{\textrm{(II)0}*}\right) P.
\end{equation}
\end{subequations}

The zeroth order problem coincides  with the  situation considered in the previous section, equations (\ref{dyn.orth.HI}), and we  can simply  write:
\begin{subequations}\label{u.0.SSCII}
\begin{eqnarray}
  v_{\vec{k}}^{\textrm{(II)}0}(\eta)=\sqrt{\frac{\hbar}{2k}}\left(-H^{\textrm{(II)}}_0\eta\right)\left(1-\frac{i}{k\eta} \right)e^{-ik\eta},\hspace{2.8cm}\quad &\textrm{for}&\quad k\neq 0, \\
   v_0^{\textrm{(II)}0}(\eta)=\sqrt{\frac{\hbar}{H^{\textrm{(II)}}_0}}\left[1-\frac{i}{6}
\left(-H^{\textrm{(II)}}_0\eta\right)^{3}\right]\left(-H^{\textrm{(II)}}_0\eta\right)^{m^2/3H^{2\textrm{(II)}}_0},\quad &\textrm{for}&\quad k= 0.
\end{eqnarray}  
\end{subequations}
Next, we consider the  first order problem.
The functions $\delta v_{\vec{k}}^{\textrm{(II)}\pm}(\eta)$ 
satisfy the dynamical equation (\ref{eq.deltau.1}), similar to that for the 
$v^{\textrm{(II)0}}_{\vec{k}\pm\vec{k}_0}(\eta)$, 
but with a source term determined by the Newtonian potential 
(which would be specified once we provide the function $P(\eta)$, something that we will  do  shortly) 
and the zeroth order functions $v^{\textrm{(II)0}}_{\vec{k}}(\eta)$ 
(given by the expressions (\ref{u.0.SSCII}) above).
It is a second order linear equation, therefore,  the solution is  univocally  determined  in terms of the 
corresponding initial data: $\delta \dot{v}_{\vec{k}}^{\textrm{(II)}\pm}(\eta_c)$ and $\delta v_{\vec{k}}^{\textrm{(II)}\pm}(\eta_c)$. 
(We take the initial time  for the SSC-II  to  be  the  collapsing time,  
$\eta_c$.)
%
The  normalization relation (\ref{orth.first})
is the only constrain on the initial data.
Two  simple  (but convenient)  choices   that  satisfy the normalization constraints are
\begin{subequations}\label{conditions}
\begin{equation}\label{conditions.1}
\delta\dot{v}_{\vec{k}}^{\textrm{(II)}\pm}(\eta_c)=0,\quad
\delta v_{\vec{k}}^{\textrm{(II)}\pm}(\eta_c)=
4v_{\vec{k}}^{\textrm{(II)0}}(\eta_c)P (\eta_c),
\end{equation}
and
\begin{equation}\label{conditions.2}
\delta\dot{v}_{\vec{k}}^{\textrm{(II)}\pm}(\eta_c)=
4\dot{v}_{\vec{k}}^{\textrm{(II)0}}(\eta_c) P (\eta_c),\quad
\delta v_{\vec{k}}^{\textrm{(II)}\pm}(\eta_c)=0.
\end{equation}
\end{subequations}
We  will pick one of them, for definiteness, the first one. As we  have indicated this  choice  completely  determines the solution (assuming the function $P (\eta)$ is  given).
As we will see next, we will not need to calculate the functions $\delta v_{\vec{k}}^{\textrm{(II)}\pm}(\eta)$ explicitly 
in order to finish our analysis.

So far, we have provided the basic ingredients  defining the   construction of  the quantum field  theory for the SSC-II, 
but
we still need to find a state $\vert\zeta^{\textrm{(II)}}\rangle\in\mathscr{H}^{\textrm{(II)}}$ such that its expectation 
value for the energy-momentum tensor leads to the desired nearly de Sitter, slightly inhomogeneous 
cosmological expansion,  characterized  by the Newtonian potential $\psi(\eta,\vec{x})$ in the space-time metric (\ref{FRW perturbed}).
The task  of  constructing that state is  rather cumbersome, and the  reader is 
encouraged to  skip the details in the first reading. The important result is that  this  construction is  carried  out in  general  and, at the  end, full  compatibility is  ensured  by  
a judicious  choice of the function $P(\eta)$ controlling the  time dependence of the Newtonian potential.

Let us first concentrate on the state. Looking at the symmetries of the space-time background it seems natural to assume that such a state  
should be of the form:
\begin{equation}\label{state.SSCII}
\vert\zeta^{\textrm{(II)}}\rangle = \ldots\vert\zeta^{\textrm{(II)}}_{-2\vec{k}_0}\rangle\otimes\vert\zeta^{\textrm{(II)}}_{-\vec{k}_0}\rangle \otimes \vert\zeta^{\textrm{(II)}}_{0} \rangle \otimes \vert\zeta^{\textrm{(II)}}_{\vec{k}_0}\rangle\otimes\vert\zeta^{\textrm{(II)}}_{2\vec{k}_0}\rangle\ldots .
\end{equation}
Here  we  are  making  an evident  abuse  of notation: The vector in Fock  space is  characterized  by parameters indicating the specific 
modes that  are excited (all  other modes  are assumed to be in the vacuum of 
the corresponding oscillator), and
 the parameters $\zeta^{\textrm{(II)}}_{\vec k}$ are  meant to indicate  exactly how  the mode   $\vec k$ has been excited. 
 We  could take these parameters to  describe, for instance, a particular coherent state for each different mode,  and then  the state   we are considering in the expression  above  would be precisely
\begin{equation}
\vert \zeta^{\textrm{(II)}}\rangle =\ldots
{\cal F}(\zeta^{\textrm{(II)}}_{-2\vec{k}_0} \hat{a}^{\textrm{(II)}\dagger}_{-2\vec{k}_0}  )
{\cal F}(\zeta^{\textrm{(II)}}_{-\vec{k}_0}\hat{a}^{\textrm{(II)}\dagger}_{-\vec{k}_0})
{\cal F}(\zeta^{\textrm{(II)}}_{0}\hat{a}^{\textrm{(II)}\dagger}_{0})
{\cal F}( \zeta^{\textrm{(II)}}_{\vec{k}_0}\hat{a}^{\textrm{(II)}\dagger}_{\vec{k}_0}) 
{\cal F}(\zeta^{\textrm{(II)}}_{2\vec{k}_0} \hat{a}^{\textrm{(II)}\dagger}_{2\vec{k}_0} )\ldots 
 \vert 0 ^{\textrm{(II)}} \rangle ,
\end{equation}
where $ {\cal F}(\hat{X})$ again   stands for the  object  
$ {\cal F}(\hat{X})  \propto\textrm{exp}(\hat{X})$.  
Needless is to say that we can  consider  similar excitations of each   mode   which are not necessarily  coherent,  but  the latter  will be sufficient for our purposes here.
  
The  expectation value of the field operator in such a state is given by
\begin{equation}\label{}
\phi_\zeta^{\textrm{(II)}}(x)=\phi^{\textrm{(II)}}_{\zeta,0}(\eta)+\left(\delta\phi^{\textrm{(II)}}_{\zeta,\vec{k}_0}(\eta)e^{i\vec{k}_0\cdot\vec{x}}+c.c.\right)
+\left(\delta\phi^{\textrm{(II)}}_{\zeta,2\vec{k}_0}(\eta)e^{i2\vec{k}_0\cdot\vec{x}}+c.c.\right) +\ldots.
\end{equation}
Contrary to what happens for the SSC-I, expression (\ref{SV.SSCI}), the excitation in each mode  leads now to
space-dependencies  with  three  characteristic  wavelengths. For  instance,  even if just   the   $k=0$ mode is  excited,  the field 
 expectation value will include   terms  with the  characteristic   behaviour $e^{\pm i\vec{k}_0\cdot\vec x}$.
In fact, for a general state of the form (\ref{state.SSCII}) we have
\begin{subequations}\label{phi.0.delta.k.0}
\begin{eqnarray}
\hspace{-.7cm}L^{3/2}\phi^{\textrm{(II)}}_{\zeta,0}(\eta) &=&\zeta^{\textrm{(II)}}_0 v_0^{\textrm{(II)0}}(\eta) 
+ \varepsilon [ \zeta_{-\vec{k}_0}\delta v^{\textrm{(II)}+}_{-\vec{k}_0}(\eta)+
 \zeta_{\vec{k}_0}\delta v^{\textrm{(II)}-}_{\vec{k}_0}(\eta)] + c.c.,\label{phi.0.SSCII}\\
\hspace{-.7cm}L^{3/2}\delta\phi^{\textrm{(II)}}_{\zeta,\vec{k}_0}(\eta)&=& 
\zeta^{\textrm{(II)}}_{\vec{k}_0}v^{\textrm{(II)0}}_{\vec{k}_0}(\eta)+\zeta^{\textrm{(II)}*}_{-\vec{k}_0}v^{\textrm{(II)0}*}_{-\vec{k}_0}(\eta)
+\varepsilon [\zeta^{\textrm{(II)}}_0\delta v^{\textrm{(II)}+}_0(\eta)+\zeta^{\textrm{(II)}*}_0\delta v^{\textrm{(II)}-*}_0 (\eta)
\nonumber\\
&& +
\zeta^{\textrm{(II)}*}_{-2\vec{k}_0}\delta v^{\textrm{(II)}+*}_{-2\vec{k}_0}(\eta)+\zeta^{\textrm{(II)}}_{2\vec{k}_0}\delta v^{\textrm{(II)}-}_{2\vec{k}_0}(\eta)
]
,\label{delta.phi}\\
\hspace{-.7cm}L^{3/2}\delta\phi^{\textrm{(II)}}_{\zeta,2\vec{k}_0}(\eta)&=& 
\zeta^{\textrm{(II)}}_{2\vec{k}_0}v^{\textrm{(II)0}}_{2\vec{k}_0}(\eta)+\zeta^{\textrm{(II)}*}_{-2\vec{k}_0}v^{\textrm{(II)0}*}_{-2\vec{k}_0}(\eta)
+\varepsilon [\zeta^{\textrm{(II)}}_{\vec{k}_0}\delta v^{\textrm{(II)}+}_{\vec{k}_0}(\eta)+\zeta^{\textrm{(II)}*}_{-{\vec{k}_0}}\delta v^{\textrm{(II)}-*}_{-{\vec{k}_0}}(\eta)
\nonumber\\
&& + 
\zeta^{\textrm{(II)}*}_{-3\vec{k}_0}\delta v^{\textrm{(II)}+*}_{-3\vec{k}_0}(\eta)+\zeta^{\textrm{(II)}}_{3\vec{k}_0}\delta v^{\textrm{(II)}-}_{3\vec{k}_0}(\eta)],
\end{eqnarray}
\end{subequations}
with similar expressions for the other $\delta\phi^{\textrm{(II)}}_{\zeta,n\vec{k}_0}(\eta)$ (for positive integers $n$).
We are considering a coherent state for all the different modes. However, 
all that  is required for the validity of the expressions above  is that $\zeta^{\textrm{(II)}}_{\pm n\vec{k}_0}\equiv \langle \hat{a}^{\textrm{(II)}}_{\pm n\vec{k}_0}\rangle$. 
For convenience,  we have considered  here a state  with a very simple form, and in particular we have assumed  there is no entanglement between the different modes in (\ref{state.SSCII}). 
Note that we could set   $\delta\phi^{\textrm{(II)}}_{\zeta,n\vec{k}_0}(\eta)=0$ for all $n\ge 2$, simply  by  
imposing  the required  relations between the parameters $\zeta^{\textrm{(II)}}_{\pm\vec{k}_0}$, $\zeta^{\textrm{(II)}}_{\pm 2\vec{k}_0}$, $\zeta^{\textrm{(II)}}_{\pm 3\vec{k}_0}$, etc.
In principle, the details  behind  such  relations could be used to determine  the  exact  values of the  undetermined parameters, 
and  it  is easy to see  that  $|\zeta^{\textrm{(II)}}_{\pm n\vec{k}_0}|\sim \varepsilon^n |\zeta^{\textrm{(II)}}_{0}|$.  
Recalling that we are  interested here just on  a first  order  in
 $ \varepsilon$ calculation, it is  clear that
all the terms in $\varepsilon \zeta_{\pm\vec{k}_0}$ and $\zeta_{\pm n\vec{k}_0}$ (with $n\ge 2$) in the expressions (\ref{phi.0.delta.k.0}) can be disregarded.
Thus,  in practice and to the order we are working here,
we can simply write $\vert\zeta^{\textrm{(II)}}\rangle = \vert\zeta^{\textrm{(II)}}_{-\vec{k}_0}\rangle \otimes \vert\zeta^{\textrm{(II)}}_{0} \rangle \otimes \vert\zeta^{\textrm{(II)}}_{\vec{k}_0}\rangle$. 
In particular that implies $|\delta\phi^{\textrm{(II)}}_{\zeta,\vec{k}_0}|\sim \varepsilon \phi^{\textrm{(II)}}_{\zeta,0}$, and, from now on, we will write 
\begin{equation}\label{descomposicion}
\phi_\zeta^{\textrm{(II)}}(x)=\phi^{\textrm{(II)}}_{\zeta,0}(\eta)
+\varepsilon\left(\delta\tilde{\phi}^{\textrm{(II)}}_{\zeta,\vec{k}_0}(\eta)e^{i\vec{k}_0\cdot\vec{x}}+c.c.\right),
\end{equation}
where we have used the  notation $\varepsilon\delta\tilde{\phi}^{\textrm{(II)}}_{\zeta,\vec{k}_0}
\equiv \delta\phi^{\textrm{(II)}}_{\zeta,\vec{k}_0} $.
Thus we  have,  up to order $\varepsilon$,
  \begin{subequations}\label{phi.0.delta.k.0.ordervarepsilon}
\begin{eqnarray}
\hspace{-.7cm}L^{3/2}\phi^{\textrm{(II)}}_{\zeta,0}(\eta) &=&\zeta^{\textrm{(II)}}_0 v_0^{\textrm{(II)0}}(\eta) 
 + c.c.,\label{phi.0.SSCII.ordervarepsilon}\\
\hspace{-.7cm}L^{3/2}\varepsilon\delta \tilde\phi^{\textrm{(II)}}_{\zeta,\vec{k}_0}(\eta)&=& 
\zeta^{\textrm{(II)}}_{\vec{k}_0}v^{\textrm{(II)0}}_{\vec{k}_0}(\eta)
+\zeta^{\textrm{(II)}*}_{-\vec{k}_0}v^{\textrm{(II)0}*}_{-\vec{k}_0}(\eta)
+ \varepsilon[\zeta^{\textrm{(II)}}_0\delta v^{\textrm{(II)}+}_0(\eta)
+\zeta^{\textrm{(II)}*}_0\delta v^{\textrm{(II)}-*}_0 (\eta)].
\label{delta.phi.ordervarepsilon}
%
%
\end{eqnarray}
\end{subequations}
The conditions above are necessary in order to ensure that  there are no  terms in $e^{\pm i n\vec{k}_0\cdot\vec{x}}$ (with $n\ge 2$) appearing in the
 expectation value of the energy-momentum tensor.  That, in turn, is necessary to  ensure  compatibility of our state ansatz with 
Einstein's  equations, and  the assumption  that
  those terms do not appear to the order $\varepsilon$ in the expression for the Newtonian potential of the SSC-II.

To the first order in $\varepsilon$ the $00$, $0i$ and $i=j$  components of Einstein's field equations take the form: 
\begin{subequations}\label{einstein.full}
\begin{eqnarray}
&& 3\mathcal{H}^{2\textrm{(II)}}-2\varepsilon \left[ \left(k_0^2 P +3\mathcal{H}^{\textrm{(II)}}\dot{P}\right) e^{i\vec{k}_0\cdot\vec{x}} +c.c. \right]= \label{einstein.full.1}\\
&& \hspace{.5cm}4\pi G \left\lbrace
(\dot{\phi}_{0,\zeta}^{\textrm{(II)}})^2+a^{2\textrm{(II)}}m^2(\phi_{0,\zeta}^{\textrm{(II)}})^2 + \right.\nonumber\\
&& \hspace{1.cm} \left. 2\varepsilon\left[\left(\dot{\phi}_{\zeta,0}^{\textrm{(II)}}\delta\dot{\tilde{\phi}}^{\textrm{(II)}}_{\zeta,\vec{k}_0}+
a^{2\textrm{(II)}}m^2\phi_{\zeta,0}^{\textrm{(II)}}\delta\tilde{\phi}^{\textrm{(II)}}_{\zeta,\vec{k}_0}+
 a^{2\textrm{(II)}}m^2(\phi_{\zeta,0}^{\textrm{(II)}})^2 P \right)e^{i\vec{k}_0\cdot\vec{x}} +c.c.\right]\right\rbrace,\nonumber\\
&& \varepsilon \left[\left(\dot{P}+\mathcal{H}^{\textrm{(II)}} P\right)e^{i\vec{k}_0\cdot\vec{x}} +c.c. \right] = 4\pi G \left\lbrace\varepsilon 
\left[\left(\dot{\phi}_{\zeta,0}^{\textrm{(II)}}\delta\tilde{\phi}^{\textrm{(II)}}_{\zeta,\vec{k}_0}\right)e^{i\vec{k}_0\cdot\vec{x}} +c.c. \right]\right\rbrace,\label{einstein.full.2}\\
&&-\mathcal{H}^{2\textrm{(II)}}-2\dot{\mathcal{H}}^{\textrm{(II)}}+2\varepsilon \left[\left(\ddot{P}+3\mathcal{H}^{\textrm{(II)}}\dot{P}+2(\mathcal{H}^{\textrm{2(II)}}+2\dot{\mathcal{H}}^{\textrm{(II)}})P\right) e^{i\vec{k}_0\cdot\vec{x}}+c.c.\right] = \label{einstein.full.3}\\
&&\hspace{.5cm}  4\pi G \left\lbrace (\dot{\phi}_{0,\zeta}^{\textrm{(II)}})^2-a^{2\textrm{(II)}}m^2(\phi_{0,\zeta}^{\textrm{(II)}})^2 +\right.\nonumber\\
&&\hspace{1cm} \left. 2\varepsilon\left[\left(
\dot{\phi}_{\zeta,0}^{\textrm{(II)}}\delta\dot{\tilde{\phi}}^{\textrm{(II)}}_{\zeta,\vec{k}_0}-
a^{\textrm{2(II)}}m^2\phi_{\zeta,0}^{\textrm{(II)}}\delta\tilde{\phi}^{\textrm{(II)}}_{\zeta,\vec{k}_0}+
 a^{2\textrm{(II)}}m^2(\phi_{\zeta,0}^{\textrm{(II)}})^2 P
-2(\dot{\phi}_{\zeta,0}^{\textrm{(II)}})^2P\right)e^{i\vec{k}_0\cdot\vec{x}}+c.c.\right] \right\rbrace.\nonumber
\end{eqnarray}
\end{subequations}
Here (\ref{einstein.full.1}) and (\ref{einstein.full.2}) are two constraints due to the diffeomorphism invariance of the theory,
and (\ref{einstein.full.3}) a dynamical equation. The $i\neq j$ equations vanish to this order.

Let us concentrate first on the space-independent 
part of the SSC-II.
Integrating the $00$ and the $i=j$ components of Einstein equations over a spatial hypersurface  of constant $\eta$ 
we obtain
\begin{subequations}\label{einstein.HI.SSCII}
 \begin{eqnarray}
  3\mathcal{H}^{2\textrm{(II)}} &=& 4\pi G\left((\dot{\phi}_{0,\zeta}^{\textrm{(II)}})^2+a^{2\textrm{(II)}}m^2(\phi_{0,\zeta}^{\textrm{(II)}})^2 \right),\label{einstein.HI.1.new}\\
  \mathcal{H}^{2\textrm{(II)}}+2\dot{\mathcal{H}}^{\textrm{(II)}}  &=& -4\pi G\left((\dot{\phi}_{0,\zeta}^{\textrm{(II)}})^2-a^{2\textrm{(II)}}m^2(\phi_{0,\zeta}^{\textrm{(II)}})^2 \right).\label{einstein.HI.2.new}
 \end{eqnarray}
\end{subequations}
Again, these are the standard Friedmann equations.
The spatial integral of the equations $0i$ and $i\neq j$ vanishes (to the first order in $\varepsilon$).
The equations (\ref{einstein.HI.SSCII}) are analogous to those obtained in Section \ref{before the collapse} for the
 SSC-I, and the state of the zero mode get fixed like in the previous case (see expression
(\ref{epsilon1.m})  and recall that, to the order we are working here, the terms in $\varepsilon \zeta_{-\vec{k}_0}$ and $\varepsilon \zeta_{\vec{k}_0}$ must be neglected in
 (\ref{phi.0.SSCII}), i.e. see expression (\ref{phi.0.SSCII.ordervarepsilon}) above).
%
Note that, in contrast  with the homogeneous and isotropic case  corresponding to the  SSC-I, in the present case the system (\ref{einstein.HI.SSCII}) 
represents just  the first contribution in a series expansion, and although here we  will  limit   ourselves to the first  order, 
it  is  clear however  that, in principle,  we  could   extend the  analysis   to any desired  order.

Introducing the equations (\ref{einstein.HI.SSCII}) into the space-dependent first order system (\ref{einstein.full}) we obtain
\begin{subequations}\label{einstein.NHI}
\begin{eqnarray}
&&\hspace{-1.2cm} k_0^2 P +3\mathcal{H}^{\textrm{(II)}}\dot{P}= -4\pi G 
\left(\dot{\phi}_{\zeta,0}^{\textrm{(II)}}\delta\dot{\tilde{\phi}}^{\textrm{(II)}}_{\zeta,\vec{k}_0}+
a^{2\textrm{(II)}}m^2\phi_{\zeta,0}^{\textrm{(II)}}\delta\tilde{\phi}^{\textrm{(II)}}_{\zeta,\vec{k}_0}+
 a^{2\textrm{(II)}}m^2(\phi_{\zeta,0}^{\textrm{(II)}})^2 P \right),\label{einstein.NHI.1}\\
&&\hspace{-1.2cm} \dot{P}+\mathcal{H}^{\textrm{(II)}} P  = 4\pi G 
\left(\dot{\phi}_{\zeta,0}^{\textrm{(II)}}\delta\tilde{\phi}^{\textrm{(II)}}_{\zeta,\vec{k}_0}\right),\label{einstein.NHI.2}\\
&&\hspace{-1.2cm} \ddot{P}+3\mathcal{H}^{\textrm{(II)}}\dot{P}+2\left(\mathcal{H}^{\textrm{2(II)}}+2\dot{\mathcal{H}}^{\textrm{(II)}}\right)P = \label{einstein.NHI.3b} \nonumber \\
&&\hspace{.1cm}   4\pi G \left(
\dot{\phi}_{\zeta,0}^{\textrm{(II)}}\delta\dot{\tilde{\phi}}^{\textrm{(II)}}_{\zeta,\vec{k}_0}-
a^{\textrm{2(II)}}m^2\phi_{\zeta,0}^{\textrm{(II)}}\delta\tilde{\phi}^{\textrm{(II)}}_{\zeta,\vec{k}_0}+
 a^{2\textrm{(II)}}m^2(\phi_{\zeta,0}^{\textrm{(II)}})^2 P
-2(\dot{\phi}_{\zeta,0}^{\textrm{(II)}})^2P\right). \label{einstein.NHI.3c}
\end{eqnarray}
\end{subequations}
The key result, and the aspect that enables us to carry out the construction in 
a complete manner is the following fact: 
the equations (\ref{einstein.NHI}) can be combined into a single dynamical 
equation for the Newtonian potential, which is independent of the matter fields  
first order quantities $\delta\tilde{\phi}^{\textrm{(II)}}_{\zeta,\vec{k}_0}$ and $\delta\dot{\tilde{\phi}}^{\textrm{(II)}}_{\zeta,\vec{k}_0}$,
\begin{equation}\label{dyn.Psi0}
 \ddot{P}+2\left[3\mathcal{H}^{\textrm{(II)}}+\frac{a^{2\textrm{(II)}}m^2\phi_{\zeta,0}^{\textrm{(II)}}}{\dot{\phi}_{\zeta,0}^{\textrm{(II)}}}
\right]\dot{P}+\left[k_0^2+2\mathcal{H}^{\textrm{2(II)}}+4\dot{\mathcal{H}}^{\textrm{(II)}}+8\pi G(\dot{\phi}_{\zeta,0}^{\textrm{(II)}})^2+
2\frac{a^{2\textrm{(II)}}m^2\phi_{\zeta,0}^{\textrm{(II)}}}{\dot{\phi}_{\zeta,0}^{\textrm{(II)}}}\mathcal{H}^{\textrm{(II)}}\right]P
 =0.
\end{equation}
In fact, we can now use Friedmann equations 
to write 
(\ref{dyn.Psi0}) with coefficients that depend on the the scale factor and its first and second time derivatives alone, 
by expressing $a^{\textrm{(II)}}m\phi_{\zeta,0}^{\textrm{(II)}}/\dot{\phi}_{\zeta,0}^{\textrm{(II)}}$ as 
$-[(2\mathcal{H}^{2\textrm{(II)}}+ 
  \dot{\mathcal{H}}^{\textrm{(II)}})/
  (\mathcal{H}^{2\textrm{(II)}}- 
  \dot{\mathcal{H}}^{\textrm{(II)}})]^{1/2}$ (we have  taken the  negative sign  because during slow-roll   $\dot{\phi}_{\zeta,0}^{\textrm{(II)}}$ and $\phi_{\zeta,0}^{\textrm{(II)}}$ must have opposite signs).
 %
We can go further by 
using the definition of the slow-roll parameter, 
$\dot{\mathcal{H}}^{\textrm{(II)}}/
 \mathcal{H}^{2\textrm{(II)}} = 1 - \epsilon^{\textrm{(II)}}$, and
express the above equation in the simpler looking way
  \begin{equation}\label{dyn.Psi.Ist order}
 \ddot{P}+2\left[ 3 \mathcal{H}^{\textrm{(II)}}-{\cal A^{\textrm{(II)}}} 
 \right]\dot{P} 
 +\left[k_0^2+
 2 (3-\epsilon^{\textrm{(II)}}) \mathcal{H}^{2\textrm{(II)}}   -
  2{\cal A^{\textrm{(II)}}} 
  \mathcal{H}^{\textrm{(II)}}
  \right] P =0 .
\end{equation}
Here we have used that 
$[(2\mathcal{H}^{2\textrm{(II)}}+ 
  \dot{\mathcal{H}}^{\textrm{(II)}})/
  (\mathcal{H}^{2\textrm{(II)}}- 
  \dot{\mathcal{H}}^{\textrm{(II)}})]^{1/2} = [(3-\epsilon^{\textrm{(II)}})/\epsilon^{\textrm{(II)}}]^{1/2}$,  
and defined  
$ {\cal A^{\textrm{(II)}}} \equiv \sqrt{3/\epsilon^{\textrm{(II)}}}ma^{\textrm{(II)}} (  1 -\epsilon^{\textrm{(II)}}/6)$,
expression valid up to the first order in the slow-roll parameter.
One might be concerned about the $1/\sqrt{\epsilon^{\textrm{(II)}}}$,
which would be extremely large during a phase of slow-roll inflation.
However, we should note that this factor appears multiplied by the scalar field  
mass $m$, and, just as in the case of the SSC-I (remember expression 
(\ref{epsilon1.m}) in Section \ref{before the collapse}), we would have 
$ H_0^{\textrm{(II)}} =  m ( 1 + r\epsilon^{\textrm{(II)}})/\sqrt{3\epsilon^{\textrm{(II)}}}$.
Although large, this is simply the natural inflationary scale. (Note that we have  
kept the higher order correction term $r$ that was not explicit in equation 
(\ref{epsilon1.m}) to be consistent with the expansion being first order  
in $ \epsilon^{\textrm{(II)}}$.)
Making use of the expression for the scale factor 
$a^{\textrm{(II)}}=\mathcal{H}^{\textrm{(II)}}/
H^{\textrm{(II)}}$, we find that the equation (\ref{dyn.Psi.Ist order}) becomes simply
  \begin{equation}\label{dyn.Psi.Ist order-2}
 \ddot{P}
 + \epsilon^{\textrm{(II)}}  (1+6r) \mathcal{H}^{\textrm{(II)}}  
 \dot{P} 
 +  \left[k_0^2
  - \epsilon^{\textrm{(II)}}  (1-6r) \mathcal{H}^{2\textrm{(II)}} 
   \right] P =0.
\end{equation}
The general solution to the equation (\ref{dyn.Psi.Ist order}) depends only on the 
zero mode (the space-independent part of the universe) 
and the 
initial conditions for the Newtonian potential, $P_c\equiv P(\eta_c)$ and 
$\dot{P}_c \equiv\dot{P}(\eta_c)$,
   \begin{equation}\label{sol.Psi.Ist order}
      P(\eta) = C_1 \,  \eta^{\frac{1}{2}[1+(6r+1)\epsilon^{\textrm{(II)}}]} J_{\alpha} (-k\eta) +  
          C_2 \,  \eta^{\frac{1}{2}[1+(6r+1)\epsilon^{\textrm{(II)}}]} Y_{\alpha} (-k\eta) ,
    \end{equation}
where $J_{\alpha}(-k\eta)$ and $Y_{\alpha}(-k\eta)$ are the Bessel functions of first
and second kind,  
$\alpha = [1 + 3 (1-2r)\epsilon^{\textrm{(II)}}]/2$ to the first order in $\epsilon^{\textrm{(II)}}$,
and $C_1$ and $C_2$ two constants that will be determined by the initial conditions.
We will not be making use of this explicit solution in the rest of the analysis. 
However, it is worth noting that this represents a damped  
oscillation of the Newtonian potential.
Once we have an expression for the function $P(\eta)$, equation (\ref{sol.Psi.Ist order}), the mode solutions determining the quantum field theory
construction become fully determined as well, as we already noted bellow equation (\ref{conditions.2}).

Regarding the initial values for the function $P(\eta)$, we note that the problem 
has a fundamental symmetry $\phi \to -\phi$, and, for definiteness, we will be 
assuming from now on that $\phi_{\zeta,0}^{\textrm{(II)}} > 0$.
 Making use of the two constraints  (\ref{einstein.NHI.1}) and (\ref{einstein.NHI.2}),  we  can 
express the initial values   that  would determine the specific  solution $P(\eta)$ in the form,\footnote{In order to study what happens when 
$k_0^2-\epsilon^{\textrm{(II)}}\mathcal{H}^{2\textrm{(II)}}=0$ (or close to that point) 
one  would need to  include  the next order in the series expansion. 
However, we are not going to analyse this here.}
\begin{equation}\label{c.i.primer}
\left( {\begin{array}{c}
 P \\
\dot{P}  \\
 \end{array} } \right) =
 \frac{\sqrt{4\pi G \epsilon^{\textrm{(II)}} }\mathcal{H}^{\textrm{(II)}}}{k_0^2- \mathcal{H}^{2\textrm{(II)}}\epsilon^{\textrm{(II)}}} 
\left( {\begin{array}{cc}
   3\mathcal{H}^{\textrm{(II)}}- {\cal A^{\textrm{(II)}}}
\; & 1\\
 - k_0^2 -  (3-\epsilon^{\textrm{(II)}})\mathcal{H}^{2\textrm{(II)}}+{\cal A^{\textrm{(II)}}}\mathcal{H}^{\textrm{(II)}} \; &  -\mathcal{H}^{\textrm{(II)}} \\
 \end{array} } \right)
 \cdot
\left( {\begin{array}{c}
 \delta\tilde\phi^{\textrm{(II)}}_{\zeta,\vec{k}_0}  \\
 \delta\dot{\tilde\phi}^{\textrm{(II)}}_{\zeta,\vec{k}_0}  \\
 \end{array} } \right).
\end{equation}
The  equations 
(\ref{c.i.primer}) apply for  $\eta\ge\eta_c$, and in particular they are valid 
at the  SSC-II   side  of collapsing time, so the system (\ref{c.i.primer}) can be used in order to infer the initial conditions for the Newtonian potential in terms of the 
characteristics of the collapse, $\delta{ \tilde\phi}^{\textrm{(II)}}_{\zeta,\vec{k}_0}(\eta_c)$ and 
$\delta\dot{ \tilde\phi}^{\textrm{(II)}}_{\zeta,\vec{k}_0}(\eta_c)$.
Given the values for $P_c$ and $\dot{P}_c$
at the collapsing time (or equivalently, $\delta \tilde\phi^{\textrm{(II)}}_{\zeta,\vec{k}_0}(\eta_c)$ and 
$\delta\dot{ \tilde\phi}^{\textrm{(II)}}_{\zeta,\vec{k}_0}(\eta_c)$), we 
have thus a completely determined space-time metric,  
and, as discussed in connection  with  (\ref{conditions.2}),
this then determines the set of normal modes for the SSC-II.
We can also use the values for $\delta \tilde\phi^{\textrm{(II)}}_{\zeta,\vec{k}_0}(\eta_c)$ and 
$\delta\dot{ \tilde\phi}^{\textrm{(II)}}_{\zeta,\vec{k}_0}(\eta_c)$, together with the expression (\ref{delta.phi}) 
and the identities (\ref{conditions.1}), to determine the value of  the parameters $\zeta^{\textrm{(II)}}_{\vec{k}_0}$ and $\zeta^{\textrm{(II)}}_{-\vec{k}_0}$ (we have
 two equations for two unknowns leading to  definite  values for the  latter).
A simple  and  naturally  expected conclusion can be seen: 
the homogeneous and isotropic part of the universe determines  the state of the zero mode $\vert\zeta^{\textrm{(II)}}_0\rangle$ (or vice versa), whereas 
the values for $P_c$ and $\dot{P}_c$ at the collapse time 
  help determine the quantum state for  the   modes $\pm \vec{k}_0$ (i.e. the ``mode   states"  $\vert\zeta^{\textrm{(II)}}_{-\vec{k}_0}\rangle$ 
and $\vert\zeta^{\textrm{(II)}}_{\vec{k}_0}\rangle$).

It is worth commenting at this point that the seemingly ``strange'' connection that 
we have found between the excitation of the mode $ \vec{k}_0$ and that of its  
higher harmonics ($ \vec{k}= n \vec{k}_0$, with $n$  natural), is intimately 
connected with the nonlinearity of the entire  proposal  
(which, as we have explained, is the general relativistic version of what occurs in  
the Shr\"odinger-Newton system), and is very similar, at the mathematical level,  
to the effect known  as ``parametric resonance'' that occurs  
in quantum optics with non-linear media \cite{Parametric1, Parametric2}.
 
This finalizes the construction of the SSC-II. 
We  have  seen  that  it is fully determined  once  given the   values  $\delta \tilde\phi^{\textrm{(II)}}_{\zeta,\vec{k}_0}(\eta_c)$ and 
$\delta\dot{ \tilde\phi}^{\textrm{(II)}}_{\zeta,\vec{k}_0}(\eta_c)$.
Next we need to study the possibility of matching this SSC to the SSC-I on the hypersurface  corresponding to the collapse time
(see Appendix \ref{app} for a discussion of the aspect of this matching 
that is connected to the gauge  issues).



\subsection{The matching at the collapse time}\label{collapse}

As we have discussed before, when trying to describe the emergence of the seeds 
of structure in our universe, we  need to consider the transition from a 
homogeneous and isotropic SSC to another one lacking 
such symmetries.
Here we will give an explicit construction for such a transition using the 
general ideas developed in Section \ref{beyond the SSC}.
As we have already stressed, we are interested in discussing the formalism,  
rather than the completely realistic and evidently quite complicated case. We
will only consider the transition from  the 
SSC-I with $\psi(\eta, \vec{x})=0$ we discussed in Section \ref{before the collapse}, 
to a situation where a single nontrivial mode $\vec{k}_0$ in the Newtonian 
potential is excited. That is, the SSC-II with $\tilde \psi_{\vec{k}}(\eta)= P(\eta) \delta_{\vec{k}\vec{k}_0}$ 
described in 
Section \ref{After the collapse}. 
 We note  that there are two  general  issues to be  treated in this context:   a)  the actual  matching conditions between 
 the SSCI and the SSCII, and  b) the  characterization of the  target state in the  Hilbert space of the SSC-I , which  in turn will be used to characterize  the  state of the SSC-II.

As indicated before we will be considering that at time $\eta_{c}$ the mode 
$\vec{k}_0$ of the state $\vert\xi^{\textrm{(I)}}\rangle$ undergoes a collapse. (Remember  
that we are attempting to formalize the description of that 
novel $-$and evidently unknown$-$ aspect of physics we have argued should  
be taking place in the early universe, 
described in this paper as the collapse of the wave function).
Following the ideas developed in Section \ref{beyond the SSC}, we assume that, first, the  characteristics  of the collapse  are 
determined  within the initial  SSC. We  will assume that  
such characterization  is  encoded  in a  state  within the 
  initial Hilbert space.  We  will  often  refer  to  such state   (using a very loose,  but heuristically helpful   language) as the state  the system 
``is tempted to 
jump into"  or the ``target  state". In our case that will be 
a state belonging to the Hilbert space $\mathscr{H}^{\textrm{(I)}}$  corresponding to the homogeneous and isotropic SSC-I,  but will not  be  {\it the state} that  
makes up the SSC-I.  In accordance  with the  comments  above, we  will  assume that  such state  corresponds to a tendency of 
excitation in the $\vec{k}_0$ mode, related in some way  to the  collapse  process that we are  about to describe, 
\begin{equation}\label{SSCI.SSCII}
\vert\xi^{\textrm{(I)}}\rangle=\vert\xi^{\textrm{(I)}}_0\rangle \;\rightarrow\; \vert\zeta^{\textrm{(I)}}\rangle_{\textrm{target}}= \vert\zeta^{\textrm{(I)}}_{-\vec{k}_0}\rangle\otimes \vert\xi^{\textrm{(I)}}_0\rangle
\otimes \vert\zeta^{\textrm{(I)}}_{\vec{k}_0}\rangle,
\end{equation}
with all the other modes remaining in their previous  state. 
We will consider how this particular target  state $\vert\zeta^{\textrm{(I)}}\rangle_{\textrm{target}}$ in 
$\mathscr{H}^{\textrm{(I)}}$ is  chosen later on. 
However,  as  we  have  anticipated, what we need to face is the fact  that the target state in 
$\mathscr{H}^{\textrm{(I)}}$, together with $g^{\textrm{(I)}}_{\mu\nu}(x)$, $\hat{\phi}^{\textrm{(I)}}(x)$ and $\hat{\pi}^{\textrm{(I)}}(x)$, 
can not represent a new SSC. 
Thus, we  need  a new SSC  corresponding to a  specific  version of  the  SSC-II, 
$\{ g^{\textrm{(II)}}_{\mu\nu}(x),\hat{\phi}^{\textrm{(II)}}(x),\hat{\pi}^{\textrm{(II)}}(x),\mathscr{H}^{\textrm{(II)}},\vert\zeta^{\textrm{(II)}}\rangle\}$, 
such that the state $\vert\zeta^{\textrm{(II)}}\rangle$ in 
$\mathscr{H}^{\textrm{(II)}}$ is in some way related to the state $\vert\zeta^{\textrm{(I)}}\rangle_{\textrm{target}}$ in $\mathscr{H}^{\textrm{(I)}}$.  
In the simple case considered  here  
  we  can focus on the operator
 $\hat{\phi}^{\textrm{(I)}}_{\vec{k}_0}(\eta)$, with $\hat{\phi}^{\textrm{(I)}}(x)=\sum_{\vec{k}}\hat{\phi}^{\textrm{(I)}}_{\vec{k}}(\eta)e^{i\vec{k}\cdot\vec{x}}$, 
and  write  $\phi_{\zeta_t,0}^{\textrm{(I)}}(\eta_c)\equiv \, _\textrm{target}\langle\zeta^{\textrm{(I)}}\vert\hat{\phi}^{\textrm{(I)}}_{0}(\eta_c)\vert\zeta^{\textrm{(I)}}\rangle_\textrm{target} $ 
 and $\varepsilon \delta\tilde\phi_{\zeta_t, \vec{k}_0}^{\textrm{(I)}}(\eta_c)\equiv \, _\textrm{target}\langle\zeta^{\textrm{(I)}}\vert\hat{\phi}^{\textrm{(I)}}_{\vec{k}_0}(\eta_c)\vert\zeta^{\textrm{(I)}}\rangle_\textrm{target}$,  even though,
 as  we have stressed,  the
target  state is  an element of  $\mathscr{H}^{\textrm{(I)}}$ but is  not part of any SSC. Here, the subindex ``$t$" in $\zeta_t$  refers to the fact that the quantity corresponds to the expectation value in 
 such a target state.
 
As it  was anticipated in Section \ref{beyond the SSC}, we will consider that the identification of the SSC-I and the SSC-II is guided by the 
expectation value of the energy-momentum tensor, expression (\ref{recipe.collapses}). We  will see this in detail in the following and, in particular, we  will see  
how  that helps  determining the state corresponding to the SSC-II  in terms of the target state.
%
%
To the zeroth order in $\varepsilon$  the  requirement (\ref{recipe.collapses}) gives:
\begin{subequations}\label{identities.collapsing}
\begin{eqnarray} 
(\dot{\phi}_{\zeta_t,0}^{\textrm{(I)}})^2+a^{\textrm{(I)}2}m^2(\phi_{\zeta_t,0}^{\textrm{(I)}})^2  &=& (\dot{\phi}_{\zeta,0}^{\textrm{(II)}})^2+a^{\textrm{(II)}2}m^2(\phi_{\zeta,0}^{\textrm{(II)}})^2 ,\label{identities.collapsing-1}\\
(\dot{\phi}_{\zeta_t,0}^{\textrm{(I)}})^2-a^{\textrm{(I)}2}m^2(\phi_{\zeta_t,0}^{\textrm{(I)}})^2  
&=& (\dot{\phi}_{\zeta,0}^{\textrm{(II)}})^2-a^{\textrm{(II)}2}m^2(\phi_{\zeta,0}^{\textrm{(II)}})^2 .\label{identities.collapsing-2}
\end{eqnarray}
\end{subequations}
From  these equations it is easy to  conclude that $ (\dot{\phi}_{\zeta_t,0}^{\textrm{(I)}})^2 =(\dot{\phi}_{\zeta,0}^{\textrm{(I)}})^2$ and $ a^{\textrm{(I)}2}(\phi_{\zeta_t,0}^{\textrm{(I)}})^2 
=a^{\textrm{(II)}2}(\phi_{\zeta,0}^{\textrm{(II)}})^2$. We  will assume that  there is no  jump in the  scale factor, and also that there is no jump in sign in the  respective field 
expectation values. Under these assumptions we are led to 
$ \dot{\phi}_{\zeta_t,0}^{\textrm{(I)}}=\dot{\phi}_{\zeta,0}^{\textrm{(I)}}$ and $ \phi_{\zeta_t,0}^{\textrm{(I)}} 
=\phi_{\zeta,0}^{\textrm{(II)}}$.

Now,  let us proceed to  discuss the matching to the first order in $\varepsilon$. At that order, expression (\ref{recipe.collapses}) gives:
\begin{subequations}\label{identities.collapsing-B}
\begin{eqnarray} 
\dot{\phi}_{\zeta_t,0}^{\textrm{(I)}}\delta\dot{ \tilde\phi}^{\textrm{(I)}}_{\zeta_t,\vec{k}_0}+
a^{\textrm{(I)}2}m^2\phi_{\zeta_t,0}^{\textrm{(I)}}\delta \tilde\phi^{\textrm{(I)}}_{\zeta_t,\vec{k}_0} &=& 
\dot{\phi}_{\zeta,0}^{\textrm{(II)}}\delta\dot{ \tilde\phi}^{\textrm{(II)}}_{\zeta,\vec{k}_0}+
a^{\textrm{(II)}2}m^2
( \phi_{\zeta,0}^{\textrm{(II)}}\delta \tilde\phi^{\textrm{(II)}}_{\zeta,\vec{k}_0}
+ (\phi_{\zeta,0}^{\textrm{(II)}})^2 P),\label{identities.collapsing-3}\\
 \dot{\phi}_{\zeta_t,0}^{\textrm{(I)}}\delta \tilde\phi^{\textrm{(I)}}_{\zeta_t,\vec{k}_0} &=&  
 \dot{\phi}_{\zeta,0}^{\textrm{(II)}}\delta \tilde\phi^{\textrm{(II)}}_{\zeta,\vec{k}_0},\label{identities.collapsing-4}\\
\dot{\phi}_{\zeta_t,0}^{\textrm{(I)}}\delta\dot{ \tilde\phi}^{\textrm{(I)}}_{\zeta_t,\vec{k}_0}-
a^{\textrm{(I)}2}m^2\phi_{\zeta_t,0}^{\textrm{(I)}}\delta \tilde\phi^{\textrm{(I)}}_{\zeta_t,\vec{k}_0} &=& \dot{\phi}_{\zeta,0}^{\textrm{(II)}}\delta\dot{ \tilde\phi}^{\textrm{(II)}}_{\zeta,\vec{k}_0}-
a^{\textrm{(II)}2}m^2
( \phi_{\zeta,0}^{\textrm{(II)}}\delta \tilde\phi^{\textrm{(II)}}_{\zeta,\vec{k}_0}
+(\phi_{\zeta,0}^{\textrm{(II)}})^2 P )
\nonumber\\
& & -2(\dot{\phi}_{\zeta,0}^{\textrm{(II)}})^2
 P.
 \label{identities.collapsing-5}
\end{eqnarray}
\end{subequations}  
First, we can use expression (\ref{identities.collapsing-4})  together  with our previous results to conclude
that $ \delta \tilde\phi^{\textrm{(I)}}_{\zeta_t,\vec{k}_0} = \delta \tilde\phi^{\textrm{(II)}}_{\zeta,\vec{k}_0}$. 
Next, subtracting equation (\ref {identities.collapsing-5}) from  (\ref{identities.collapsing-3}) and 
using again the  previous  results we  find that
 $(\dot{\phi}_{\zeta,0}^{\textrm{(II)}})^2P 
 =0$. Thus, as we  will  assume  that after the collapse the universe remains 
in a slow-roll expansion, $\dot{\phi}_{\zeta,0}^{\textrm{(II)}}\not=0$, we can conclude that $ P
 =0$. This might seem problematic, but let us recall that the matching conditions are supposed to hold only 
 at the time of collapse, $\eta= \eta_c$, thus  $ P(\eta_c)=0$, but $P$ at  later times  need not  vanish.
  On the other hand, as we will see, this  will drastically  simplify  our  analysis. 
%
%
Finally, using these results and adding equations (\ref {identities.collapsing-5}) and (\ref{identities.collapsing-3}) we 
find that $\delta\dot{ \tilde\phi}^{\textrm{(I)}}_{\zeta_t,\vec{k}_0}
=\delta\dot{ \tilde\phi}^{\textrm{(II)}}_{\zeta,\vec{k}_0}$.

It is  worth mentioning that  equations  (\ref{identities.collapsing-1}) and (\ref{identities.collapsing-2}) have been 
obtained from $(T_{00}^{\textrm{(I)}})_{\zeta_t,0}=(T_{00}^{\textrm{(II)}})_{\zeta,0}$ and $(T_{ii}^{\textrm{(I)}})_{\zeta_t,0}=(T_{ii}^{\textrm{(II)}})_{\zeta,0}$, 
whereas (\ref{identities.collapsing-3}), (\ref{identities.collapsing-4}) and (\ref{identities.collapsing-5}) from $(\delta T_{00}^{\textrm{(I)}})_{\zeta_t,0}=(\delta T_{00}^{\textrm{(II)}})_{\zeta,0}$, $(\delta T_{0i}^{\textrm{(I)}})_{\zeta_t,0}=(\delta T_{0i}^{\textrm{(II)}})_{\zeta,0}$ and $(\delta T_{ii}^{\textrm{(I)}})_{\zeta_t,0}=(\delta T_{ii}^{\textrm{(II)}})_{\zeta,0}$.
For a scalar field, and  up to the first order in $\varepsilon$,
the other components of the energy-momentum tensor do not contain additional information.
%
%
%
%
%
Of course the identities (\ref{identities.collapsing}) and (\ref{identities.collapsing-B}) are only 
valid at the collapsing time.  What is more, the target state only plays a  role at that particular time. 


Recapitulating, we will ask that at the matching $a^{\textrm{(II)}}(\eta_c)= a^{\textrm{(I)}}(\eta_c)$, and  found 
%
\begin{subequations}\label{id. collapse}
\begin{eqnarray}
 \phi_{\zeta,0}^{\textrm{(II)}}(\eta_c)=\phi_{\zeta_t,0}^{\textrm{(I)}}(\eta_c),& \quad &
 \dot{\phi}_{\zeta,0}^{\textrm{(II)}}(\eta_c)=\dot{\phi}_{\zeta_t,0}^{\textrm{(I)}}(\eta_c),\label{id. collapse.1}\\
 \delta \tilde\phi^{\textrm{(II)}}_{\zeta,\vec{k}_0}(\eta_c)=\delta \tilde\phi^{\textrm{(I)}}_{\zeta_t,\vec{k}_0}(\eta_c), & \quad &
 \delta\dot{ \tilde\phi}^{\textrm{(II)}}_{\zeta,\vec{k}_0}(\eta_c)=\delta\dot{ \tilde\phi}^{\textrm{(I)}}_{\zeta_t,\vec{k}_0}(\eta_c),
\end{eqnarray}
\end{subequations}
and $ P
(\eta_c)=0$.
We can use now  expressions (\ref{einstein.HI.SSCII}) to conclude  that $H^{\textrm{(I)}}=H^{\textrm{(II)}}$ and $\epsilon^{\textrm{(I)}}=\epsilon^{\textrm{(II)}}$, i.e.  we have  continuity   
for the space-independent part of the space-time background (the  spatial metric and  extrinsic  curvature).

Within each SSC, the  expectation  values of field and momentum operators satisfy the Ehrenfest theorem 
(recall that within  each SSC there is nothing  exotic  going on, and each  mode  of the field  is  essentially a harmonic 
oscillator). This can be used to compute the quantities  
$\phi_{\zeta,0}^{\textrm{(II)}}(\eta_c)$ and $\delta \tilde\phi^{\textrm{(II)}}_{\zeta,\vec{k}_0}(\eta_c)$ (see the expression 
(\ref{phi.0.delta.k.0}) above and remember that, to the order we are working here, 
the parameters $\varepsilon \zeta_{\pm\vec{k}_0}$ and $\zeta_{\pm n\vec{k}_0}$ 
(with $n\ge 2$)
can be disregarded, as we did for instance in equation (\ref{phi.0.delta.k.0.ordervarepsilon})).
The values for $\dot{\phi}_{\zeta,0}^{\textrm{(II)}}(\eta_c)$ and $\delta\dot{ \tilde\phi}^{\textrm{(II)}}_{\zeta,\vec{k}_0}(\eta_c)$ can be obtained
from the time derivatives of the expressions given in (\ref{phi.0.delta.k.0}).
However, note  that  $\phi_{\zeta_t,0}^{\textrm{(I)}}(\eta_c)$ and $\delta \tilde\phi^{\textrm{(I)}}_{\zeta_t,\vec{k}_0}(\eta_c)$   
correspond  to expectation values in  a  target  state,  which,  
as we have  been  emphasising, despite  being an element of the Hilbert space of the SSC-I,
is not {\it the state}  characterizing 
 the  SSC-I. Thus, in  general, such quantities  can  exhibit  spatial dependences.   
 In fact, these are given by the expressions 
\begin{subequations}\label{targetual.phi.delta.phi}
\begin{eqnarray}
L^{3/2}\phi_{\zeta_t,0}^{\textrm{(I)}}(\eta_c)&=&\xi^{\textrm{(I)}}_0 u_0^{\textrm{(I)}}(\eta_c)+c.c. ,\label{HI-2}\\
L^{3/2} \varepsilon\delta\tilde \phi_{\zeta_t, \vec{k}_0}^{\textrm{(I)}}(\eta_c)&=& 
\zeta^{\textrm{(I)}}_{\vec{k}_0}u^{\textrm{(I)}}_{\vec{k}_0}(\eta_c)+\zeta^{\textrm{(I)}*}_{-\vec{k}_0}u^{\textrm{(I)}*}_{\vec{k}_0}(\eta_c), \label{inh-2}
\end{eqnarray}
\end{subequations}
and $\dot{\phi}_{\zeta_t,0}^{\textrm{(I)}}(\eta_c)$ and $\delta\dot{ \tilde \phi}^{\textrm{(I)}}_{\zeta_t,\vec{k}_0}(\eta_c)$
by their time derivatives.
%
%
Comparing the identities (\ref{id. collapse}) with the expressions (\ref{phi.0.delta.k.0.ordervarepsilon}) and (\ref{targetual.phi.delta.phi}), we arrive to $\zeta_0^{\textrm{(II)}}=\xi_0^{\textrm{(I)}}$, 
$\zeta_{-\vec{k}_0}^{\textrm{(II)}}=\zeta_{-\vec{k}_0}^{\textrm{(I)}}$ and $\zeta_{\vec{k}_0}^{\textrm{(II)}}=\zeta_{\vec{k}_0}^{\textrm{(I)}}$ (recall that,
 according to the choice given in (\ref{conditions.1}), for the case $P (\eta_c)=0$ we have $\delta\dot{v}_{\vec{k}}^{\textrm{(II)}\pm}(\eta_c)=\delta v_{\vec{k}}^{\textrm{(II)}\pm}(\eta_c)=0$). That is, using  the mode solutions  chosen  in Sections \ref{before the collapse} and \ref{After the collapse},
the parameters $\zeta_{\vec{k}}^{\textrm{(II)}}$ characterizing the state of the 
inflaton field in the SSC-II can be directly read from the parameters $\zeta_{\vec{k}}^{\textrm{(I)}}$ characterizing the target state in $\mathscr{H}^{\textrm{(I)}}$. 
Thus,  once  the target state is determined,  the  complete  SSC-II gets fixed, because  as we showed in  Section 
\ref{After the collapse} everything is  determined  there once  we specify the  SSC state.

From $P(\eta_c)=0$ and the equation (\ref{c.i.primer}) we find that, 
at the time of collapse, we should have
\begin{equation}\label{constraint.collapse}
(3\mathcal{H}^{\textrm{(II)}}-\mathcal{A}^{\textrm{(II)}})\, \delta \tilde\phi^{\textrm{(II)}}_{\zeta,\vec{k}_0}(\eta_c)
+\delta\dot{ \tilde\phi}^{\textrm{(II)}}_{\zeta,\vec{k}_0}(\eta_c) 
=0.
\end{equation} 
Combining (\ref{c.i.primer}) and (\ref{constraint.collapse}) we obtain the explicit  
expression for ${\dot P}(\eta_c)$, which shows that, even  though  the Newtonian potential is continuous at $\eta_c$, its time derivative is not,
\begin{equation}\label{psi.dot}
{\dot P} (\eta_c) =
 \sqrt{4\pi G\epsilon^{\textrm{(II)}}}\mathcal{H}^{\textrm{(II)}}
\delta \tilde\phi^{\textrm{(II)}}_{\zeta,\vec{k}_0}(\eta_c).
\end{equation}
This jump in the time derivative of the Newtonian potential at the collapsing time gives rise to 
the primordial perturbation in the $\vec{k}_0$ mode.
That is, the  spatial metric  is continuous at the transition, but its time derivative is not, i.e. at the collapsing time we will have a continuous but  non-smooth 
description for the space-time manifold. 

Finally, 
we must  give a prescription for the election of the target state $\vert\zeta^{\textrm{(I)}}\rangle_{\textrm{target}}$ involved in
 the collapse, expression (\ref{SSCI.SSCII}).
As it was anticipated in Section \ref{beyond the SSC}, and as it has been usual in our previous treatments of the subject, we will 
consider that the target state is chosen stochastically, guided by the quantum uncertainties, at the time of collapse, of some field operators
 evaluated in the  pre-collapse state $\vert\xi^{\textrm{(I)}}\rangle$.  
These  operators have been usually taken to be the corresponding modes of the inflaton field or their conjugate momenta, or some  
combination thereof, and sometimes even both. We  had  rather large freedom  in what  we chose in that regard.
However, 
in the present  analysis, we find that the 
field and momentum
can not be assumed  to change their  expectation value   during the collapse in an  arbitrary way:   The condition (\ref{recipe.collapses}) imposes $P_c=0$, and then the relation (\ref{constraint.collapse}) 
above. Here, we will consider that the collapse is guided by the quantum uncertainties associated to the modes of the field operator, 
and that the  immediate post-collapse   expectation value  of momentum  operator  is  such that  the condition $P_c=0$  is satisfied. 

We need to clarify one more point before analysing the determination of the target state. As discussed in  \cite{Perez:2005gh},  
in order for the collapse to resemble as much as possible the implementation of the reduction postulate (which is connected with 
Hermitian observables), we decompose the operators $\hat{\phi}^{\textrm{(I)}}_{\vec{k}}(\eta)$ in its real and imaginary parts, 
$\hat{\phi}^{\textrm{(I)}}_{\vec{k}}(\eta)=\hat{\phi}^{\textrm{(I)R}}_{\vec{k}}(\eta)+i\hat{\phi}^{\textrm{(I)I}}_{\vec{k}}(\eta)$,  and focus  the  collapse on those.
The  operators  we must consider are then 
\begin{equation}
\hat{\phi}^{\textrm{(I)R,I}}_{\vec{k}}(\eta)=\frac{1}{\sqrt{2}}\left(u^{\textrm{(I)}}_{\vec{k}}(\eta)\hat{a}^{\textrm{(I)R,I}}_{\vec{k}}+u^{\textrm{(I)}*}_{\vec{k}}(\eta)\hat{a}^{\textrm{(I)R,I}\dagger}_{\vec{k}}\right)
\end{equation}
and
\begin{equation}
\hat{a}^{\textrm{(I)R}}_{\vec{k}}=\frac{1}{\sqrt{2}}\left(\hat{a}^{\textrm{(I)}}_{\vec{k}}+\hat{a}^{\textrm{(I)}}_{-\vec{k}}\right), \quad
\hat{a}^{\textrm{(I)I}}_{\vec{k}}=\frac{-i}{\sqrt{2}}\left(\hat{a}^{\textrm{(I)}}_{\vec{k}}-\hat{a}^{\textrm{(I)}}_{-\vec{k}}\right).
\end{equation}
With these definitions $\hat{\phi}^{\textrm{(I)R,I}}_{\vec{k}}(\eta)$ are
Hermitian operators (i.e. $\hat{\phi}^{\textrm{(I)R,I}}_{\vec{k}}(\eta)=\hat{\phi}^{\textrm{(I)R,I}\dagger}_{\vec{k}}(\eta)$), but the commutation relations between $\hat{a}^{\textrm{(I)R}}_{\vec{k}}$ and $\hat{a}^{\textrm{(I)I}}_{\vec{k}}$
 are non-standard,
\begin{equation}\label{nonstandard.commutators}
[\hat{a}^{\textrm{(I)R}}_{\vec{k}},\hat{a}^{\textrm{(I)R}\dagger}_{\vec{k}'}]= (\delta_{\vec{k},\vec{k}'}+\delta_{\vec{k},-\vec{k}'}),\quad
[\hat{a}^{\textrm{(I)I}}_{\vec{k}},\hat{a}^{\textrm{(I)I}\dagger}_{\vec{k}'}]= (\delta_{\vec{k},\vec{k}'}-\delta_{\vec{k},-\vec{k}'}),
\end{equation}
with all the other commutators vanishing. 

Now we turn to specifying the target state in SSC-I, and thus the relevant state of SSC-II.   As discussed  in  \cite{Perez:2005gh}, we will  be assuming, in a loose analogy with standard quantum mechanics, 
that the collapse is somehow similar to an imprecise measurement of the operators $\hat{\phi}^{\textrm{(I)R,I}}_{\vec{k}_0}(\eta)$, and that 
the final results will be guided by 
\begin{equation}\label{uncer.collapse}
\varepsilon \delta\tilde \phi^{\textrm{(II)R,I}}_{\zeta_t,\vec{k}_0}(\eta_c) = x_{\vec{k}_0}^{\textrm{R,I}}
\sqrt{\langle 0^{\textrm{(I)}}_{\vec{k}_0}\vert\left[\Delta\hat{\phi}^{\textrm{(I)}}_{\vec{k}_0}(\eta_c)\right]^2\vert 0^{\textrm{(I)}}_{\vec{k}_0} \rangle}=  x_{\vec{k}_0}^{\textrm{R,I}}\sqrt{\frac{ 1}{2}}\left|v_{\vec{k}_0}^{\textrm{(I)}}(\eta_c)\right| ,
\end{equation}
with $x_{\vec{k}_0}^{\textrm{R,I}}$ taken to be two independent  random variables distributed according to a
 Gaussian function centred at zero with unit-spread. The expression (\ref{uncer.collapse}), 
 together with the relation (\ref{constraint.collapse}), determines the values for $\varepsilon \delta\tilde \phi^{\textrm{(I)}}_{\zeta_t,\vec{k}_0}(\eta_c)$ and $\varepsilon \delta\dot{\tilde\phi}^{\textrm{(I)}}_{\zeta_t,\vec{k}_0}(\eta_c)$ at the
  collapsing time (in terms of the random variables $x_{\vec{k}_0}^{\textrm{R,I}}$), and then the state $\vert\zeta^{\textrm{(I)}}\rangle_{\textrm{target}}$, and  thus $ \vert\zeta^{\textrm{(II)}}\rangle$
\begin{subequations}\label{zeta.collapse}
\begin{eqnarray}
\zeta_{\vec{k}_0}^{\textrm{(II)}}=\zeta_{\vec{k}_0}^{\textrm{(I)}} &=& -i\hbar^{-1}a^2(\eta_c)\varepsilon
\left[\dot{v}_{\vec{k}_0}^{\textrm{(I)}*}(\eta_c)\delta\tilde\phi_{\zeta_t,\vec{k}_0}^{\textrm{(I)}}(\eta_c)-
v_{\vec{k}_0}^{\textrm{(I)}*}(\eta_c)\delta\dot{\tilde\phi}_{\zeta_t,\vec{k}_0}^{\textrm{(I)}}(\eta_c)\right],\\
\zeta_{-\vec{k}_0}^{\textrm{(II)}}=\zeta_{-\vec{k}_0}^{\textrm{(I)}} &=& -i\hbar^{-1}a^2(\eta_c)\varepsilon
\left[\dot{v}_{\vec{k}_0}^{\textrm{(I)}*}(\eta_c)\delta\tilde\phi_{\zeta_t,\vec{k}_0}^{\textrm{(I)}*}(\eta_c)-
v_{\vec{k}_0}^{\textrm{(I)}*}(\eta_c)\delta\dot{\tilde\phi}_{\zeta_t,\vec{k}_0}^{\textrm{(I)}*}(\eta_c)\right].
\end{eqnarray}
\end{subequations}
That is, the state of the inflaton field in the SSC-II will be given by
$\vert\zeta^{\textrm{(II)}}\rangle= \vert\zeta^{\textrm{(II)}}_{-\vec{k}_0}\rangle\otimes \vert\zeta^{\textrm{(II)}}_0\rangle
\otimes \vert\zeta^{\textrm{(II)}}_{\vec{k}_0}\rangle$, with $\zeta^{\textrm{(II)}}_0=\xi^{\textrm{(I)}}_0$, 
 and $\zeta_{\vec{k}_0}^{\textrm{(II)}}$ and $\zeta_{-\vec{k}_0}^{\textrm{(II)}}$ determined in terms of the values for 
$\varepsilon\delta\tilde\phi^{\textrm{(I)}}_{\zeta_t,\vec{k}_0}(\eta_c)$ and 
$\varepsilon\delta\dot{\tilde\phi}^{\textrm{(I)}}_{\zeta_t,\vec{k}_0}(\eta_c)$ at the collapsing time, 
expressions (\ref{zeta.collapse}). Regarding the metric tensor, its space-independent part will not 
be affected by the collapse, $H^{\textrm{(I)}}=H^{\textrm{(II)}}$ and $\epsilon^{\textrm{(I)}}=\epsilon^{\textrm{(II)}}$. On the other hand, the 
Newtonian potential will be given by  the solution of   the  differential equation expression (\ref{dyn.Psi0}), with initial conditions $P_c
=0$ and $\dot{P}_c
$ determined
 in terms of the values for $\varepsilon\delta\tilde\phi^{\textrm{(I)}}_{\zeta_t,\vec{k}_0}(\eta_c)$ and $\varepsilon\delta\dot{\tilde\phi}^{\textrm{(I)}}_{\zeta_t,\vec{k}_0}(\eta_c)$ by (\ref{c.i.primer}) (see also the expression (\ref{psi.dot}) above). 
 
One should avoid  being deceived  by the close  relationship  between  the  target state $\vert\zeta^{\textrm{(I)}}\rangle_{\textrm{target}}$ and  
the state $ \vert\zeta^{\textrm{(II)}}\rangle$.  While,  as  we have seen, their  corresponding expectation values for the basic field   
operators at  the  collapse time  are  the same,  their subsequent  
evolution will in general deviate  from one  another.
 In  particular, 
the Newtonian potential  will become non-vanishing, and thus the  modes of the  SSC-II construction will involve non-vanishing 
$ \delta v_{\vec{k}}$'s (see  equation (\ref{ansatz}) above). These aspects  might become  relevant in 
  a detailed  analysis of the resulting   primordial spectrum,  something that  lies  well  beyond the  scope  of the present  manuscript.
    
We emphasize that we have found that, although the Newtonian potential is 
continuous at the collapse time, its time derivative is not. This means we do 
not really have a true space-time description of the process.  
In the next section we will briefly discuss our views on this problematic  
aspect of our results, and in future  works we hope to investigate ways in which 
this aspect of our formalism might be improved.


\section{Discussion}\label{conclusions}

We have considered the generic joint description of gravitation in interaction with a quantum field, to the extent that this can 
be done without a fully workable theory of quantum gravity. 
Our proposal is formalized in terms of what we call a Semiclassical 
Self-consistent Configuration (SSC), which is nothing but a combination of quantum field theory on a background space-time, with the 
requirement that the state of the matter fields and the space-time geometry  
be connected through the semiclassical Einstein equations.   
We have applied this approach to a simple inflationary cosmological model, describing both a perfectly homogeneous and isotropic
configuration, and a situation that deviates from the former one by a slight 
excitation of a particular inhomogeneous and anisotropic perturbation. 
We have considered in detail a proposal for describing the  ``space-time'' 
where 
such a perturbation  actually {\it emerges} from the initial homogeneous and
isotropic configuration
as the result of the collapse of the wave function of the inflaton 
field.  To our knowledge, this represents  the first time  such a detailed description  
of the process of  emergence of  
structure is  ever presented.   
We believe that the treatment developed in this paper can be useful in uncovering  
how the collapse of a wave function can be made compatible with a  fundamental theory of gravity.
If and when we would be in possession of a fully workable and satisfactory quantum  theory of  gravitation, and  are  able to describe in detail its 
semiclassical regime, we should be able to explore the exact behaviour of the  
gravitational degrees of freedom on the collapse hypersurface.
However, even in the absence of such a theory, a study of these  issues could produce interesting insights  into some  of the  features   that  such theory   should  contain.
We have argued that the general formalism developed in this work should  
be useful in detailed studies of the various theories  
involving wave function collapse  
(such as \cite{Bassi:2003gd, Karolyhazy1966, Diosi1984, Diosi1987, Diosi1989,GRW1,Pearle1989}), and
in particular in their applications to situations where the gravitational 
back-reaction becomes important, as well as in proposals such as the stochastic  
gravity of Hu and Verdaguer \cite{Stoch} (see the Appendix \ref{B} for more 
details). 
    
Regarding the inflationary regime (which provided the motivation for the   
development of this formalism), it is clear that what we have done here  
is just a starting point, as we have limited ourselves to the study of a single 
collapse. In order to analyse the problem of the emergence of the seeds of cosmic 
structure in a complete fashion we  would need to consider multiplicity of 
collapses, occurring in multiple times and  involving  all the different modes, as  we have  done, schematically, in  previous  works
 \cite{Perez:2005gh, DeUnanue:2008fw, Leon:2010fi, Leon:2010wv}. 
  However, in contrast  with what was done there, the  study  can now, in principle, be  carried  out  using a  well defined  and precise   formalism   as presented  here.
  That formalism would allow  us to consider issues such as the degree  
to which  ``energy conservation'' is violated  during a collapse, 
  and possibly to analyse  its  effects on   the   evolution of  the universe 
during and immediately after inflation. Moreover, as  discussed  in  section \ref{After the collapse},  
  the  formalism  suggests  the  existence of correlations between the excitation level  of the modes   $\vec k$ and that of  its higher harmonics, a feature that  is reminiscent of  the  so  called  ``parametric  resonances" 
  occurring in  quantum optics with  nonlinear  materials \cite{Parametric1, Parametric2}.

It is worthwhile to contrast the formalism we have developed in this  work with the standard treatment of inflation,
which is, at the fundamental level, essentially perturbative.
Such a treatment is based on the separation of a background 
(involving both the space-time metric and the inflaton field), which is described  
at the classical level, reserving the quantum treatment just for the linear perturbations. 
The formalism we have considered here is, in principle, amenable to a non-perturbative treatment, 
even though in practice we are often impeded from carrying that out simply because of the usual limitations that 
prevent a general non-perturbative treatment of a quantum field theory.  
Nonetheless, if one manages to overcome this limitation on the quantum field side, for instance  
through treatments based on lattice approaches, or simply by considering some quantum field solvable model, 
the scheme could be appropriate to the consistent inclusion of gravity.    
One obvious advantage of such a setup is that the path to considering higher order perturbations (i.e. anything beyond first order  
perturbations) is clear and well defined from the beginning. On the other hand, it is quite evident that this cannot be considered as a  
fully satisfactory description of nature because of all the well known arguments indicating that we need a quantum theory of gravitation, 
see for instance references \cite{de Witt1967, Wheler1968, Carlip2008}.
Nevertheless, it seems reasonable to assume that a theory like this can be suitable for a description of a situation where the measures  
and  estimates (i.e. classical and quantum mechanical) of the space-time curvature are well below the Planck regime, 
and where the matter fields energy-momentum tensor have uncertainties that are ``not too large" (i.e.  see Penrose's arguments).  
We will be  working under the  assumption  that this would be the case for most of the inflationary regime we are interested in, and for the post-inflationary 
cosmological regimes that follow it.
  
However, as we have previously argued, this cannot be the full story if we want to be able to account for the transition from the completely 
homogeneous and isotropic universe which is usually associated with the early and mid stages of inflation (and that we have considered   
in Section \ref{before the collapse} within a very  simple model described by the SSC-I), to the late situation where inhomogeneities and anisotropies  
are present (and  which was described in a simplified manner involving just one excited mode by the SSC-II of Section \ref{After the collapse}).   
Accounting for this phenomena requires a departure from the established  
unitary 
evolution in the form of  some sort of ``collapse of the wave function" 
(as  considered  by Diosi \cite{Diosi1984, Diosi1987, Diosi1989}, Ghirardi, Rimini and Weber \cite{GRW1}, or Pearle \cite{Pearle1989}; see also \cite{Bassi:2003gd}), 
a feature that might presumably find its full justification in a deeper theory of quantum gravity, as it has been previously 
discussed by R. Penrose \cite{Penrose:1996cv, Penrose2005}. 
In the Appendix \ref{A} we have presented a speculation of how something like 
that might be tied to the   
resolution of the problem of time in quantum gravity, based on  the findings  of \cite{Gambini:2006ph} that the usage 
of a physical clock can naturally introduce effective deviations from unitary evolution.
   
Nonetheless, regardless of such speculations, the issue we need to face is how to modify the SSC formalism to include such a transition. 
Here, we have considered an attempt to do so, which involves the selection of a state within the Hilbert space 
of the SSC-I (called the ``target state'') to which the system  ``wants to jump" (to which it would jump if the result was also a SSC), and then
 finding  a new SSC for which the associated  state had, on the collapse hypersurface (a space-like hypersurface taken for simplicity 
to coincide with a homogeneous and isotropic one of the cosmological model described  by the SSC-I), the same 
energy-momentum tensor as the target state. We  can argue that this ``matching  recipe" is more or less natural, although clearly has several 
aspects that can be considered rather ad hoc. 
One can certainly consider other possible recipes.  However, as it turns out  
{\it a posteriori}, this option has some nice features, in the sense  of  limiting the degree of arbitrariness  on the specification of
 both the target state (through the requirement that the condition (\ref{constraint.collapse}) is  satisfied) and the SSC-II (in the sense 
 of fixing the expectation values of the basic field operators, and the values for the Newtonian potential and its first time derivative at
  the collapsing time, see  equation (\ref{id. collapse})).  The resulting 
``space-time" turns out to be described by a continuous 
  but {\it not} smooth metric (there is a jump in the extrinsic curvature on the 
matching hypersurface), and, as such,  the result is not truly a space-time.
In fact, we already knew that a jump associated with a quantum collapse would 
imply that since, at the  corresponding space-time points, Einstein's 
 semiclassical equations  would  not hold, simply because of the fact that for any smooth space-time the Bianchi identity implies the vanishing of $\nabla_\mu G^{\mu\nu}$,  
while  the  expectation value of the energy momentum tensor would quite generally not be divergence-less during the jump of the quantum state \cite{Perez:2005gh}.
This is certainly an unappealing aspect of the formalism, and this is one of the reasons for which
 we can not take this as a complete and satisfactory description of the problem at hand.

There are  some other aspects of the proposal that seem unsatisfactory, with one of the most problematic being the issue of general covariance. 
It is evident that the association of 
a collapse with a particular space-like hypersurface brings up the very issue 
that is  usually cause of grave concern
regarding  the compatibility with special relativity of any theory involving an 
instantaneous reduction of the wave function.
At this point, we should mention a related (but different) issue of gauge dependence
 (or independence) of the proposal. As this issue  has  led to considerable 
confusion we  have  devoted  the  Appendix \ref{app} and  turn the reader  to it for a careful discussion.
 Turning  back to the  character of  the collapse  hypersurface,  
the explicit  analysis we have  presented here ascribes to such particular space-like hypersurface a very  particular {\it physical role}:
 It separates the space-time region which is perfectly homogeneous and isotropic from the one where a particular
 kind of anisotropy has set in. It is then clear that the selection of such hypersurface is not a simple gauge choice, but it corresponds
  to part of the characterization of the proposal regarding the collapse process itself.

On the other hand, the evident tension between the existence of space-like 
hypersurfaces with particular physical properties and 
relativistic ideas is clearly a very serious issue, as it seems to imply the physical breakdown of cherished and well established principles, and 
  as  such, the problem can certainly be grave.
  Of course, this was not unexpected, as one 
  of the most troublesome aspects of the notion ``collapse of the wave function" is precisely that it seems intrinsically associated with a global and  
  instantaneous  
  process, and at this point we can only hope that  the
idea might be eventually reconciled with the  principles of relativity.
This issue is in a sense close to the core of the famous EPR {\it  gedankenexperiment}, and of course the  modern developments  
including its experimental realization.
The problem is thus not hopeless by any means, but, of course, the discussion of alternatives is well beyond the scope of the present
 paper.\footnote{We should mention here the ideas of Rovelli regarding  a relational wave
  function \cite{Rovelli1996}, the proposals designed to make the collapse models applicable to field theories, 
  which assume that the collapse is triggered at a single event and then propagates on the 
  past light cone \cite{Hellwing1970}, or the  analysis in  \cite{Myrvold2002}.}  It is nevertheless worth 
  mentioning that our view in this regard is that the novel physical process is associated and triggered by global 
  conditions, and as  such it might be both instantaneous and covariantly  defined,\footnote{Relativity  requires   the laws of  physics
    to  be  the same  in  any reference frame, but that of course does not  prevent  the existence of special frames  associated  with   a  particular  
  state of a physical system. Say a  body determines a  spacial  frame  in which  it is  at rest.} but of  course it would need to have 
   features that prevent it  from being used (and here the important word is {\it used}, which implies the possibility
    of external manipulation by conscious beings) to send information faster than light.
It is worth noting that there are various works suggesting that non-locality might 
play an even more fundamental role in a complete theory than that 
it plays in standard quantum theory (see for instance \cite{Popescu, Hamber, Giddings, Deser}).
We should, however, keep in mind that this  is  meant  only as  an effective description  of limited validity,  
and that a  truly developed  theory of quantum gravity can  naturally be   expected to be needed  in order to obtain something  
completely satisfactory.  It  is  perhaps also  worth pointing out that the issue of   compatibility with  Lorentz  invariance  is  a 
difficulty that  seems to  emerge often in connection  with bringing together the quantum  theory and   gravitation,  such as in  loop 
quantum gravity \cite{gambini, Alfaro1999, Morales2002, Alfaro2005, Alfaro20052} (similar problems  appear also  in the  so 
called   Liouville  approach to string theory 
\cite{Ellis2000, Ellis2002}), and that  in  those  cases, it  often provides important  constraints that have not  been  dealt  with  in 
full yet \cite{Collins2004, Polshinski2011}.
   
Perhaps, a simple analogy would be helpful in order to convey what is precisely what we have in mind.
Let us imagine for a moment that we  do not have a mathematical description of  curved  surfaces  (say,  for concreteness,  2-surfaces  embedded in 
3-dimensional Euclidean space), and that we only know  how to characterize  planes. Let us assume that we want to describe a certain smooth rock. 
Clearly we would not be able to do that unless the rock was completely flat. However, we can obtain what would be for most 
purposes a reasonable description of the rock by wishing a large number of tangent planes and indicating that the rock is what lies within  
the volume that the planes define (adding all other relevant information about where certain planes would end, etc).  
It is clear that at the intersection of the planes we would have certain singular behaviour (at those points we would not have a  
unique normal characterizing the rock),  but we should not be surprised by that. We know that although the sharp vertexes   
associated with the intersections are not to be taken seriously, many other features (like  for instance the  volume of the rock) 
can be reasonably expected to be well characterized with our rough description. Moreover, one might even hope  
that perhaps such limited description could be the starting point motivating the development of the differential geometry of smooth surfaces.  
It is our hope that by attempting to push the well developed theories of physics (quantum field theory on curved space-time and  
classical general relativity) to the limit in an attempt to describe the situations of interest in inflationary cosmology (which is, 
as we already noted, the only regime where both quantum theory and gravitation come together in dealing with situations  that are 
observationally accessible to us) can serve not only to push further our understanding of the specific situation, but  
it would be also useful in some manner to continuing investigations on the quantum gravity realm. It seems clear to us  
that if the collapse of the wave function is a required modification of the quantum theory brought by the quantum gravitational effects as 
it is suggested by  Doisi and Penrose, this would be the case.  
 
\acknowledgments

The work of ADT is supported by a UNAM postdoctoral fellowship and the CONACYT  grant No 101712. The work of DS is  supported  in part by the   CONACYT grant
No 101712, by  the PAPIIT-UNAM grant  IN107412-3, and  by  sabbatical fellowships  from CONACYT and DGAPA-UNAM.  DS  thanks  the  IAFE-UBA  for the  hospitality  during the sabbatical stay.

\appendix

\section{The collapse within our current understanding of physical theory}\label{A}

Let us try to frame our proposal, even if only schematically, within the  general  
current understanding of physical theory.
The  basic  idea  underlying our current considerations,  and  which was initially proposed  in \cite{Sudarsky:2009za},  
 is to connect the  problem at hand to that   encountered  when trying to write  a theory of
   quantum gravity through the canonical  quantization procedure. 
 As it is well  known,  when  following  approaches of this  kind, such as  the old  
 Wheeler de Witt proposal \cite{de Witt1967}, or its  more  modern incarnation in the form of loop  quantum gravity \cite{Rovelli},  one  ends up 
  with an atemporal theory.  This is  known as ``the problem of time in 
quantum gravity" \cite{Isham:1992ms}. That is,  in   both  schemes  one  starts  with a formulation  in which the basic   canonical variables  
describe the  geometry of a 3-spatial  hypersurface  $\Sigma$, and   characterize  the  embedding of this  3-surface in a 4-dimensional  space-time. 
 From those  quantities  one is  led to the identification of  a set of canonical variables, which we will 
 denote here generically by $({\cal G},  \Pi )$.
(In the Wheeler de Witt case this stands for the spatial metric $h_{\mu\nu}$ and a certain
  function of the extrinsic curvature $K_{\mu\nu}$, i.e. $(h_{\mu\nu}, K_{\mu\nu})$, while in loop quantum gravity
   these will be the densitized triad $E^{\mu}_i $ and connection $A_{\mu}^i $ variables, i.e. $(E^{\mu}_i , A_{\mu}^i)$).
The problem is that time, or its general relativistic  counterpart (a time function usually specified by the lapse  function and the shift vector)
simply  disappears  from the theory, given that
the Hamiltonian vanishes  when acting on the  physical states  (those  satisfying the diffeomorphism and  Hamiltonian  constraints).

The problem is then how  would one  recover a  space-time  description of  our world,  clearly an essential element  
one  would need in order  to be  able to connect the theory with observations.
One of the  most  favoured  approaches towards  addressing this  problem
is to  consider,  simultaneously  with the  geometry, some  matter fields, which  we   will  describe  here  schematically by a collection  of ordered  pairs of
  canonically variables,
 \begin{equation}
  \lbrace(\varphi_1,\pi _1),\ldots, (\varphi_n, \pi_n)\rbrace ,
  \end{equation}
 and to  identify an appropriate  variable (or combination of variables)  in the  joint  matter  gravity theory  
that  could  act  as a physical  clock $ T(\varphi_i,\pi _i, {\cal G}, \Pi)$. The  next step consists of  characterizing  the  state  for
 the remaining  variables in terms of the correlations 
of  their values  with those of the physical clock. 
That is,  one  starts  from  the   wave  function  for the  configuration variables of the theory 
$  \Phi(\varphi_1,\ldots,\varphi_n, {\cal G})$, which  must   satisfy the   so called Hamiltonian  and  momentum constraints
 $ H_\mu   \Phi(\varphi_1,\ldots\varphi_n, {\cal G} ) =0$.  Next,  one  needs to  obtain  an effective  wave function
  $\Psi$ for the remaining  variables  by  projecting $\Phi$ into  the  subspace  
  where  the operator $T(\varphi_i,\pi _i, {\cal G},\Pi)$  takes a certain range of values. 
   That is, let us denote  by $P_{T,  [t  ,  t+\delta t]} $  the projector operator  onto the subspace corresponding to the 
    region   between  $t $ and  $ t+\delta t$ of the spectrum of the operator $T$. One  then attempts  to  
    recover a Schr\"odinger-like  evolution  equation by   studying the dependence of  $ \Psi (t)  \equiv P_{T,  [t  ,  t+\delta t]} \Phi $ on the 
     parameter  $t$.
    (In the  general  relativistic setting  this   would be a  global time function  ${\cal T} $ defined  on  the reconstructed  approximate space-time).
After  obtaining, by the above  procedure, a wave function associated  with  the   spectrum of the operator $T $  we  might use  it  to  compute
  the  expectation values of the   3-dimensional geometrical operators  (say the  triad  $E^{\mu}_i(x) $ and  connection $A_{\mu}^i (x)$ variables of Ashtekar, or
    some  appropriate smoothing thereof) for  the wave function $ \Psi (t)$. Such collection of  quantities  could be  seen as providing the  
    geometrical descriptions of the  ``average"  space-time  in terms of the  $3+1$  decomposition. 
  In other words,  one  would   have  constructed  a  space-time  where the slicing  would  correspond to the   hypersurfaces  
  on which   the  geometrical    quantities   are given  by the expectations  of the projected  wave  functions $ \Psi (t)$, and  thus  
  one  would  be able  to characterize the   space-time  and  its slicing in terms of   the   lapse and shift  functions.

The precise realization of this procedure  depends  strongly on the situation and  specific  theory  for the matter  fields  
one  is  considering,  and  such  study is  quite  beyond  the scope  of the present paper, among other reasons  because  
we  do not have at this point a satisfactory and  workable  theory of quantum gravity. On the other hand,  several works  
along these  lines exist in the literature \cite{Gambini:2004pe}. 
The point we want to make here,  however, is that  in such  a setting,  the standard  Schr\"odinger equation  
emerges only as an effective description, and it is only approximately valid.  Under these  circumstances, small
  modifications to that equation would  not  be  unexpected.  In fact, in a recent analysis \cite{Gambini:2006ph} of a
quantum mechanical system, it was found  that describing it  in terms of the time measured  by  a physical clock,  rather than  an idealized  one, implied   
modifications  representing  departures  from the quantum mechanical unitary evolution.
We  consider  that  this could  be the grounds where a modification of  the Schr\"odinger evolution, involving  something akin to a collapse 
of the wave function, might find  its ultimate explanation.  There  are,  indeed,  several proposals  for such a modification centering on the 
analysis  of  standard  laboratory  situations \cite{Bassi:2003gd, Karolyhazy1966, Diosi1984, Diosi1987, Diosi1989, GRW1, Pearle1989}.  
Moreover, it  should  be noted that  with   a paradigm where the  quantum  jumps occur   generically and spontaneously, rather than  being thought as 
triggered by the decisions of observers to measure  particular  quantities,  one might  avoid  the kind of problem discussed 
in \cite{Sorkin:1993gg} (see also the idea of the objective wave function 
reduction developed by R. Penrose in Chapter 29 of reference \cite{Penrose2005}).

\section{Conection with other approaches}\label{B}

Let us compare 
the general formalism introduced in Section \ref{Effective}
with the stochastic gravity proposal by Hu and Verdaguer \cite{Stoch}.
The idea behind that approach to semiclassical gravity is to attempt to 
take into account the ``fluctuating part of the energy 
momentum tensor" of the (quantum)  mater fields through the introduction of  a  
stochastic field $\chi_{\mu\nu}(x) $. 
The proposal involves a modified version of Einstein equations written  as:
\begin{equation}
\label{Stoch}
  G_{\mu\nu}(x)   =8\pi G  (\langle \hat{T}_{\mu\nu}(x)  \rangle  +\chi_{\mu\nu}(x) ).
  \end{equation}
The proposal then assumes that the ensemble statistics  
of the stochastic term are characterized by a certain measure of 
the uncertainties of the energy momentum tensor.
  
Let us show that one of the   instantaneous  collapses,  occurring  say at $t=t_c$,  could be seen as  corresponding in the  above  scheme  to a particular contribution to the   stochastic  field at that  particular  time, the time of collapse.
  As usual in  quantum field theory  we   use the  Heisemberg picture. 
However we  assume that the state of the field is not constant  in time, 
but that as a result of the collapse process (which we will be considering  
to occur instantaneously), it ``jumps".   
Thus, the state of the field is described by
   $|\psi (t)  \rangle = \theta(t_c -t)| \xi \rangle   + \theta(t-t_c) |\zeta \rangle$,  
where $\theta(.)$ is the step function 
(it is $0$ when the argument is negative and $1$ when it is positive).
Einstein equations would then be given by  
  \begin{equation}
\label{Stoch-Collapse}
  G_{\mu\nu} (x)  =8\pi G  \langle \psi (t) |  \hat{T}_{\mu\nu}(x) |\psi (t)    \rangle  =  8\pi G ( \langle \xi  |  \hat{T}_{\mu\nu} |\xi    \rangle  +\chi_{\mu\nu}) ,
  \end{equation}
where the stochastic term reflecting the ``jump" takes the form $ \chi_{\mu\nu}\equiv \theta(t-t_c) (\langle \zeta  |  \hat{T}_{\mu\nu} |\zeta    \rangle - \langle \xi  |  \hat{T}_{\mu\nu} |\xi    \rangle )$. 
  As we  have  indicated the exact  relationship  between the  formalism  
developed  in this work and  the  different approaches  is  outside of the scope of the present paper,  but the above  considerations   indicate  the  existence  of  a relatively  close  connection.
In future works we will explore such connections  more closely in order to, 
on the one hand, use the formalism to better understand those proposals, 
and on the other hand to consider the possibility of incorporating the  
inflationary issue  that has  motivated this and previous works  within the context of such theories and proposals.

\section {Gauge conditions, change of variables and all that}\label{app}

In mathematical physics space-time is characterized by a differential 
manifold $M$ with a type $(0,2)$ tensor field $g$ defined on it.
This characterization is independent of the coordinates.  
On the same manifold we can have other fields (of 
scalar, vector or tensor nature), 
denoted generically by  
$\psi$ and representing  matter. 
Again such characterization is independent of the choice of coordinates.
Modification of coordinate choices can never mix up different fields. 
When we choose some specific coordinates, $(x^\mu)$, the metric can be written 
in their components, using for instance the basis of one forms   
naturally associated with that coordinate chart, 
$g =g_{\mu \nu}  dx^\mu dx^\nu$. This is  
a tensor of type $(0,2)$ built with tensor products of two one-forms. 
A change of coordinates will change the components $g_{\mu \nu}$. However,  
the metric as a mathematical object will not change.

Now let us see when and how the issue of ``gauge" appears, 
and how it often leads to confusion. Consider a situation where
we have two space-times with metric and matter fields defined on 
them, $(M, g, \psi)$  and $(\tilde M, \tilde g, \tilde \psi)$, and assume  
we want to compare one with the other.  
For that we need to use some diffeomorphism (which will only exist when the two differential manifolds are diffeomorphic), 
$F:M \to \tilde M$, mapping one manifold to the other. 
Then one might want to consider the differences
in  metric and fields by looking at, say, $\delta g \equiv  \tilde{g} - F^*(g)$  and $\delta \psi \equiv \tilde{\psi}- F^*(\psi) $.
When doing this, the result will evidently depend on the 
choice of $F$. This is what is often done when considering perturbations 
from, say, a homogeneous and isotropic 
space-time to one that deviates from the former by a small amount. 
This is the setting often used to considerations involving inflationary cosmology.  
There  $(M, g, \psi)$ is taken as the ``homogeneous and isotropic background'', and 
$(\tilde M, \tilde g, \tilde \psi)$ the situation representing somehow our universe. 
This is not an issue of coordinates, but it can be confused with one.  
The fact is, however, that the issue is customarily solved by  
choosing a ``gauge'' that effectively uses the symmetries of the first  
space-time to determine (to the desired perturbative order) the diffeomorphism $F$. 
In an alternative approach, which is often employed in cosmology, one considers 
suitable combinations of certain components of $ \delta g$  and $\delta \psi$ 
(associated with suitable coordinate choices),
looking for combinations that are invariant under ``small changes 
of $F$'': these are the so called  ``gauge invariant perturbations". 

The approach based on the use of gauge invariant quantities can be quite useful 
in some calculations, but often can  
make things a bit more difficult when discussing interpretational aspects. 
We can see this by noting that what is observed using our satellites is often 
described using coordinates, such as the angular coordinates on the celestial 
sphere to characterize the CMB, just to give one example. 
Moreover, the fact that in our treatment the matter fields and the space-time 
metric are so clearly distinct forces us to avoid a gauge invariant treatment 
and work with the approach based on 
fixing the gauge. 
In this paper we have chosen to work in the so called ``Newtonian'' (sometimes
also known as ``longitudinal'' or ``conformal-Newtonian'') gauge, introduced for instance
in Chapter 9.2 of reference \cite{Mukhanov2005}.

In the particular situation we want to consider in this paper we have
a scalar field living on a ``space-time''  
which is the result of gluing together two pieces.  
The first piece is the region characterized by $\eta<\eta_c$ of the SSC-I, i.e  
a perfectly homogeneous and isotropic space-time  
with the scalar field in a state where only the zero mode 
(which is also homogeneous and isotropic) is excited. 
This is described in detail in Section \ref{before the collapse}.   
The second piece is the region $\eta>\eta_c$ of the construction corresponding 
to a slightly inhomogeneous and anisotropic space-time, with
a characteristic  wave  vector $\vec k_0$ and a scalar field in a state  
where there is a nontrivial excitation not only of the zero mode, but also of several
other modes $\vec k_0$, $2\vec k_0$, $3\vec k_0$, $\dots$. This  
is described in detail in Section \ref{After the collapse}.  
The two pieces are glued together at the hypersurface $ \Sigma_c$, corresponding  
to the regions with coordinates $\eta= \eta_c$ in each of the two pieces. 
The matching of the two pieces makes up a space-time describing the emergence (in 
the traditional sense of the word: i.e. something that was  
not there ``at a given time'' is there ``at a latter time'') of 
perturbations. This is described in Section \ref{collapse}.
Regarding the coordinates, we consider the whole manifold to be covered  
by a single coordinate chart, 
($\eta, x^1, x^2, x^3$). 
The construction we have given describes something that is almost a space-time 
(i.e is a space-time except for the fact that the  extrinsic curvature is not 
continuous in the hypersurface $ \Sigma_c$), and is, 
in the mathematical sense, analogous to the formalism employed in considering  
infinitely thin matter shells (see for instance \cite{Israel}), and are thus not  
realistic in the very same sense. We expect that in a more realistic  
description the shells would have a finite thickness, and that the collapse 
hypersurface would perhaps have some small (but finite)
temporal extent. The similarity breaks down in the fact that, in the case of 
the thin shells, we have a workable and well defined theory capable of treating 
the problem to any desired degree of accuracy. 
In the  situation at hand, the collapse is expected to be described by some 
theory which does not exit yet, and as we have argued, such theory  
would probably trace its origin to the quantum gravity regime. 
Moreover, as we have discussed in Section \ref{Effective}, 
we would expect that the characterization of any collapse theory in terms of 
space-time language would only be achievable once the fundamental degrees of 
freedom for the gravitational interaction have been given an approximate classical description.
We note here that if we wanted to change coordinates, the hypersurface $ \Sigma_c$  
would in general be described in the new choice by some complicated  
function characterizing for instance its ``time coordinate'' in terms of  
the other three coordinates. Thus our ``almost" space-time would be the same  
but described in a more complicated form. The construction has been  
carried out in a specific gauge, but the result is gauge independent.  
This is essentially the same when 
making some analysis in general relativity using some  
quite specific coordinates, say studying the perihelion of Mercury using 
Schwarzschild coordinates, one needs not worry that the result is coordinate  
dependent.  
    
There is, on the other hand, a very different issue that might be confused with 
that of gauge freedom. It is connected to the question: What is the physics that  
triggers the collapse, and how does that mechanism determine that the surface 
where it should occur is the hypersurface $\eta= \eta_c$?
We of course do not know the answer, simply because we do not have a well developed and 
workable theory of quantum gravity, and much less a theory of collapse.  
The present work is only the first step in the development of a well defined 
formalism that we hope could be useful in  obtaining well defied but parametrized
predictions that might be compared with observations (as done for intense in \cite{DeUnanue:2008fw}), 
and as a result would help us learn something about the physics of collapse. 
The only thing we can say at this point is that, if something like the value of  
the uncertainties in the quantum state of the matter fields 
is connected with triggering of collapse (as it would be for instance in a 
theory of collapse based on Penrose's ideas), then the fact that the region  
described by the SSC-I is completely homogeneous leads us to conclude that these  
local uncertainties would reach the same level exactly at the same value of $\eta$, 
and thus it would be natural to expect that the collapse would be associated  
with the corresponding hypersurface. 
But this is of course just ``educated" speculation in the absence of a detailed  
theory of collapse. 

Another source of confusion comes form the use of conformal transformations and 
the subsequent mixing of variables.
One might be concerned with the fact that it seems always possible to move part of 
the degrees of freedom from the metric to the matter fields (and back) 
through a conformal transformation, and therefore the split between what 
`should' and `should not' be treated at the quantum level would be completely  
artificial, and thus intrinsically arbitrary \cite{Wald  private communication}. 
(Given a space-time metric $g$, one can introduce  
a new metric $\bar{g}$ and a new scalar field $\chi$ related to the former 
metric by $g = \chi^2 \bar{g}$,  
and then regard $\chi$ as a matter field. 
That is, one might be in doubt as to which of the two 
metrics one should describe at classical level.)
We do not share such point of view, simply because it is based on 
a classical treatment, where indeed such conformal transformations are well defined 
and meaningful. Our view is that at a truly fundamental $-$and thus quantum$-$ 
level, the space-time degrees of freedom are of a different nature than those of 
the matter degrees of freedom, and that at such level 
there would be no ambiguity whatsoever.
Of course, at the practical level we must work without a 
satisfactory theory of quantum gravity, and the ambiguity would have to be 
resolved in some other way. We will take the view that the  
resolution comes simply from considering the physical space-time metric to be 
the one for which the corresponding geodesics are associated with the paths of the 
free particles, i.e. 
the metric to which the other fields are coupled in the minimal way.


\end{document}